\documentclass[useAMS,usenatbib,referee]{biom}
\usepackage{amsmath}
\usepackage{amssymb}
\usepackage{mathtools}
\usepackage[flushleft]{threeparttable}
\usepackage{booktabs}

\usepackage{bm}
\newcommand{\bb}{\boldsymbol}
\newcommand{\T}{\mathsf{T}}

\usepackage{enumitem}
\usepackage{comment}
\usepackage{xcolor}

\usepackage{amsfonts}
\usepackage{tabularx}
\usepackage{subcaption}
\usepackage{setspace}
\usepackage{multirow}
\usepackage{comment}
\usepackage{rotating}

\usepackage[pdfencoding=auto, bookmarks=true, hidelinks]{hyperref}
\usepackage{hyperref, cleveref}
\hypersetup{
     colorlinks   = true,
     citecolor    = blue,
     linkcolor    = blue
}

\title[Leveraging Two-Phase Data to Improve Survival Predictions in Nasopharyngeal Cancer]{Leveraging Two-Phase Data for Improved Prediction of Survival Outcomes with Application to Nasopharyngeal Cancer}

\author{Eun Jeong Oh$^{1,2}$,
   Seungjun Ahn$^{3,4}$, 
   Tristan Tham$^{5}$, and
   Min Qian$^{6}$ \\
   $^{1}$Northwell, New Hyde Park, NY \\
   $^{2}$Institute of Health System Science, Feinstein Institutes for Medical Research, Manhasset, NY \\
   $^{3}$Institute for Healthcare Delivery Science, Department of Population Health Science and Policy, \\ Icahn School of Medicine at Mount Sinai, New York, NY \\
   $^{4}$Tisch Cancer Institute, Icahn School of Medicine at Mount Sinai, New York, NY \\
   $^{5}$Department of Otolaryngology - Head and Neck Surgery, Stanford University School of Medicine, \\ Palo Alto, CA \\
   $^{6}$Department of Biostatistics, Mailman School of Public Health at Columbia University, New York, NY}

\begin{document}


\pagerange{000--000} \pubyear{0000}

\volume{00}
\artmonth{0000}
\doi{xxxx}


\label{firstpage}


\begin{abstract}
Accurate survival predicting models are essential for improving targeted cancer therapies and clinical care among cancer patients. In this article, we investigate and develop a method to improve predictions of survival in cancer by leveraging two-phase data with expert knowledge and prognostic index. Our work is motivated by two-phase data in nasopharyngeal cancer (NPC), where traditional covariates are readily available for all subjects, but the primary viral factor, Human Papillomavirus (HPV), is substantially missing. To address this challenge, we propose an expert-guided method that incorporates prognostic index based on the observed covariates and clinical importance of key factors. The proposed method makes efficient use of available data, not simply discarding patients with unknown HPV status. We apply the proposed method and evaluate it against other existing approaches through a series of simulation studies and real data example of NPC patients. Under various settings, the proposed method consistently outperforms competing methods in terms of c-index, calibration slope, and integrated Brier score. 
By efficiently leveraging two-phase data, the model provides a more accurate and reliable predictive ability of survival models. \\
\end{abstract}


\maketitle

\section{Introduction}
\label{sec:intro}

Human papillomavirus (HPV) has been recognized as an important prognostic indicator in head and neck cancer, particularly oropharyngeal cancer (OPC) \citep{ang2010}. {\color{black} The importance of HPV as a prognostic factor in OPC was found to be significant enough to update the prognostic staging system for OPC to include HPV status, within the latest American Joint Committee on Cancer (AJCC) 8th Edition Cancer Staging System \citep{lydiatt2017head}.
However, the prognostic role of HPV in other head and neck sites, such as within the nasopharynx, have not been as well studied.} 
Both the nasopharynx and oropharynx subsites are anatomically contiguous within Waldeyer’s Ring, which is a ring of lymphoid-rich tissue located in the upper aerodigestive tract. Sites within Waldeyer’s ring are postulated to serve as ideal sites for HPV-driven carcinogenesis \citep{maxwell2010, huang2022}. In this context, several studies \citep{verma2018, wotman2019,  jiang2016, wu2021} have attempted to investigate the prognostic role of HPV in nasopharyngeal carcinoma (NPC). 
However, a common problem in recent studies on characterizing HPV-induced NPC \citep{verma2018, wotman2019} is the substantial proportion of unknown HPV status. \citet{wotman2019} utilized the Surveillance, Epidemiology and End Results (SEER) database in which 70\% of the NPC patients had unknown HPV status. Similarly, in the National Cancer Data Base (NCDB) study by \citet{verma2018}, 11,126 patients had unknown HPV status out of a total of 12,389 NPC patients, indicating a nearly 90\% rate of missing HPV status.

Although it is possible to analyze the patients with known HPV status only, such an analysis is likely to cause biased estimation and reduce statistical efficiency \citep{seaman2013}. Alternatively, an imputation method can be applied to handle unknown HPV status. However, while there is no recognized threshold for an acceptable percentage of missing data in imputation approaches, extra caution is warranted given such a high proportion of unknown HPV status in NPC studies.
Among different imputation methods, multiple imputation (MI) \citep{rubin2004} has been commonly used in various applications, yet the challenge of consistent variable selection across multiply imputed datasets complicates model interpretation and inference. Several authors \citep{chen2013, wood2008, wan2015, yang2005} have proposed to tackle {\color{black} the issue of consistent variable selection with MI. Furthermore, a few studies \citep{white2009imputing, bartlett2015multiple} have proposed the MI method for survival outcomes; however,} there is a lack of methodology and easy-to-use software tools {\color{black} that address consistent variable selection across multiply imputed datasets in survival settings.}

The missing pattern largely coincides with ``two-phase'' data where a majority of variables are readily available for all subjects with a few covariates of interest being missing. This happens when all patients information on traditional risk factors are gathered, but expensive biomarkers could only be collected on a much smaller cohort. 
A few studies have considered the use of survival models when data is collected based on a two-phase design \citep{twophase1, twophase2, twophase3}. 
There is, however, scarce literature on efficient use of the two-phase data for survival outcomes with the goal of improving discrimination and calibration. 


To address the aforementioned issue, we propose an expert-guided method for the two-phase data. Our goal is to enhance not only discriminatory ability but also calibration of a survival prediction model by efficiently leveraging data, rather than discarding patients with missing data. The proposed method is articulated through a two-stage procedure designed to improve the prediction of survival models. Specifically, we first develop a partially penalized Cox regression model using the full sample and observed variables (e.g., traditional covariates), while allowing covariates identified as crucial based on domain knowledge to be unpenalized. Next, we apply the model obtained from the first stage to the target samples (i.e., a cohort of patients with observed HPV status) to generate predicted risk scores for each individual. These predicted risk scores are referred to as the prognostic index. Finally, we fit the Cox model on the target samples using the prognostic index and key factors (e.g., HPV status) as covariates. This step integrates key expensive biomarkers while adjusting for the prognostic index, which represents a summary measure of an individual’s risk of the event based on the observed covariates. 

The prognostic index has been widely adopted for assessing the calibration of validated prognostic models \citep{pi1, pi2, vanhoulingen}.
The use of prognostic index also has a close connection to a transfer learning approach which aims to borrow information across different data sources. Recently, \citet{li2023} adopted transfer learning in the time-to-event outcome framework to transfer knowledge from the source cohort (i.e., large cancer registries) to the target cohort (i.e., a single cancer center). However, such an approach assumes that the same set of variables are available for both cohorts, and thus, is not applicable to our motivating example. 
In contrast, our proposed method allows the prognostic index to be re-calibrated along with the key factors for further adjustment to the target samples. Re-calibration is a statistical process allows to adapt a risk function to a different population, aiming to eliminate the over- or under-estimation of risk in the importing population \citep{harrell2001, royston2013}.
In the extreme scenario where the missingness of HPV status does not depend on any other information (i.e., missing completely at random) and the HPV status itself is not correlated with observed covariates, the estimated slope for the prognostic index will be close to 1, and thus the re-calibrated slope will remain similar. In other cases, including when data is missing at random, the re-calibrated slope will be shifted due to systematic differences by the missing data mechanism (if applicable) and collinearity between the prognostic index and key factors (if present).

The rest of this paper is organized as follows. We briefly present the Cox model and describe the proposed expert-guided method for the two-phase data. We then apply the proposed approach and compare with other existing alternatives through simulation studies and a real data example of NPC patients using the SEER database. In the end, conclusions and discussion are provided.

\section{Methods}

\subsection{Cox regression model and penalization}

Assume we observe data $(T, \delta, \bb\Phi)$ from $n$ patients, where $T$ is the observed time, $\delta$ is the censoring indicator, and $\bb{\Phi} \in \mathbb{R}^J$ is the {\color{black} $J$-dimensional} covariates or predictors. In the proportional hazards model, also known as the Cox model \citep{cox1, cox2}, the hazard function $h(t)$ is expressed as
\begin{align}
h(t) =  h_0 (t) \exp( \bb \beta^\T \bb \Phi),
\label{regular.cox}
\end{align}
where $h_0(t)$ is an unspecified non-negative baseline hazard function and $\bb \beta \in \mathbb{R}^J$ is the parameter vector. The corresponding Cox's partial likelihood is
\[
L(\bb \beta) = \prod_{r \in D} \frac{\exp( \bb \beta^\T \Phi_r)}{ \sum_{i \in R_r} \exp( \bb \beta^\T \Phi_i)},
\]
where $D$ is the index set of death times and $R_r$ is the index set of patients at risk for death at time $t_r$. The model is assumed to be sparse; that is, a partial set of the $\bb\beta$ contains zero regression coefficients. Then a natural goal is to correctly identify the set of relevant (nonzero) variables. 

Penalized methods adopt a shrinkage penalty to combine variable selection with parameter fitting. Denote $l(\bb\beta) = \log L(\bb\beta)$. The penalized estimates of $\bb\beta$ are obtained by minimizing the following objective function
\begin{align*}
- l (\bb\beta) + \text{pen}_\lambda (\bb\beta),
\end{align*}
where $\text{pen}_\lambda (\bb\beta)$ is the penalty function which depends on a tuning parameter $\lambda$. There are plenty of choices for the penalty function, including Lasso \citep{lasso}, adaptive Lasso \citep{alasso}, elastic net \citep{enet}, SCAD \citep{scad}, and Dantzig selector \citep{dantzig}.

\subsection{Expert-guided method for two-phase data}
\label{subsec:method}

For the two-phase data, let $\bb \Phi = (\bb U, \bb V)$ and $\bb\beta = (\bb\beta_U, \bb\beta_V)$, where $\bb U \in \mathbb{R}^{p}$ denotes the {\color{black} $p$-dimensional} traditional variables that are mostly observable and $\bb V \in \mathbb{R}^{d}$ denotes a few expensive {\color{black} $d$-dimensinoal} biomarkers which may only be measured on a subsample of patients $n' \ll n$. 
Usually, the dimension of $\bb U$ is moderate or high, while $\bb V$ is low-dimensional. In the example of NPC data, $\bb V$ is HPV status which is only observed for a small subset of cohort, and $\bb U$ contains baseline covariates, including age, gender, marital status, and so forth. 
In the following, we introduce an expert-guided method to efficiently handle the two-phase data. The procedure is as follows:

\begin{enumerate}

    \item[I.] In the initial stage, we develop a penalized Cox regression model on the full samples of size $n$ using $\bb U$. To reflect domain knowledge in the variable selection process, we decompose $\bb U$ into two parts, $\bb Z$ and $\bb X$, such that $\bb U = (\bb Z, \bb X)$, where $\bb Z \in \mathbb{R}^q$ contains a few key variables that one wishes to retain and $\bb X \in \mathbb{R}^{p-q}$ contains candidate predictors that are considered for variable selection in a data-driven manner. The hazard function that we consider is
    \begin{align}
        h(t) = h_0(t) \exp (\bb\eta^\T \bb Z + \bb\gamma^\T \bb X),
        \label{EG.cox}
    \end{align}
    where $(\bb \eta, \bb \gamma) \in \mathbb{R}^p$ is the parameter vector.
    To simultaneously keep $\bb Z$ in the model and perform variable selection on $\bb X$, we impose a penalty on {\color{black} $\bm \gamma$} only.
    Specifically, we estimate $(\bb\eta, \bb\gamma)$ using the partially penalized Cox regression model with adaptive Lasso which minimizes
    \begin{align}
        - l ( \bb \eta, \bb \gamma) + \lambda_n \sum_{j=1}^{p} w_j|\gamma_j|,
        \label{EG.cox2}
    \end{align}
    where $\lambda_n$ is a tuning parameter which controls model complexity and $\bb w = (w_1, \ldots, w_p)$ is a vector of weights that are used to adjust a level of penalization on individual covariates. The weights are constructed by $\widehat{\bb w} = |\widetilde{\bb\gamma}|^{-\delta}$ for some $\delta > 0$, where $\widetilde{\bb\gamma}$ is a root-($n$/$p$)-consistent estimator. The 5- or 10-fold cross-validation can be used to select an optimal pair of $(\delta, \lambda_n)$.
    
    \item[II.] Next, the model obtained in the initial stage is applied to the target samples of size $n'$ to generate the prognostic index $\widehat{\zeta} = \widehat{\bb\eta}^\T \bb Z + \widehat{\bb\gamma}^\T \bb X \in \mathbb{R}^{1}$ for each individual. We then fit the Cox model on the target samples using $\widehat{\zeta}$ and $\bb V$ as covariates. That is, we consider
    \begin{align}
        h(t) = h_0(t) \exp(\theta_0 \widehat{\zeta} + \bb\theta_1^\T \bb V),
        \label{EG.cox3}
    \end{align}
    where the parameter vector $(\theta_0, \bb\theta_1) \in \mathbb{R}^{J-p+1}$ is estimated by maximizing the corresponding Cox's partial likelihood.
\end{enumerate}

\begin{figure}[!ht] 
\centering
\includegraphics[scale=.44]{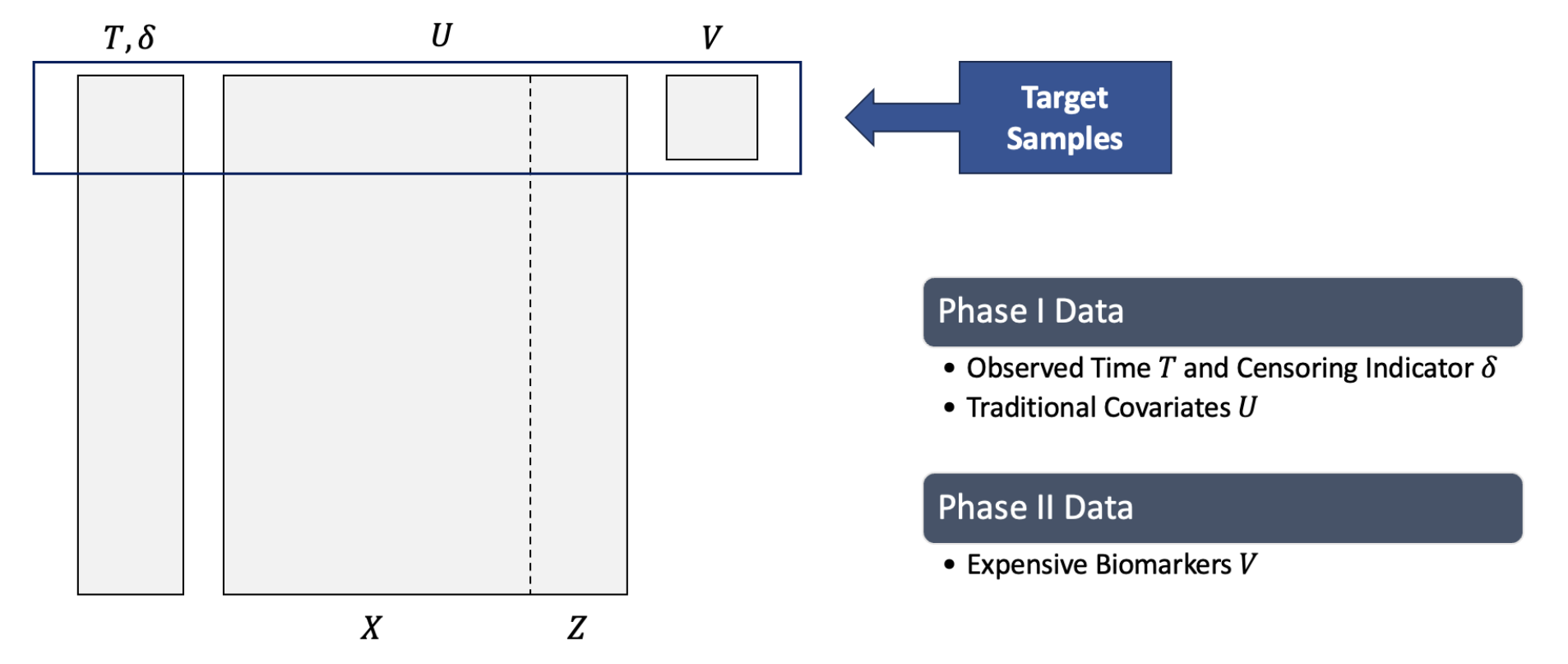}
\caption{Overview of two-phase data linked to the proposed expert-guided method.}
\label{fig:basic}
\end{figure}

The expert-guided method for two-phase data is a two-stage procedure, contrasting with complete case analysis that ignores patients with observed $\bb U$ and missing $\bb V$. 
We consider adaptive Lasso for the penalization in \eqref{EG.cox2} as it ensures that the set of non-zero coefficients is correctly identified with probability converging to one, and the estimated coefficients within this set are asymptotically normal \citep{alasso}.
Different procedures have been proposed for constructing adaptive Lasso weights in
the literature, including univariable regression \citep{univregest1, univregest2} in a low-dimensional setting and a ridge regression \citep{alasso, mult2} or a preliminary regular Lasso \citep{prelimlasso1, prelimlasso2, prelimlasso3} in a high-dimensional setting. However, no consensus has been reached. 
In the present article, we use perturbed elastic net estimates \citep{zouzhang2009} for adaptive Lasso weights. {\color{black} For implementation of the proposed method, we use the R package \texttt{glmnet}  (Friedman et al., 2010) in the first stage and \texttt{survival} (Therneau, 2023) in the second stage.}
Figure \ref{fig:basic} below provides an overview of two-phase data and its connection to the proposed expert-guided method. {\color{black} The underlying assumptions regarding the missing mechanism are discussed in Web Appendix A.}

\noindent \textbf{Remarks.} 
\begin{enumerate}

    \item[1.] The prognostic index is a linear combination of selected predictors among $\bb U$ weighted by the estimated regression coefficients. It is essentially the log of relative hazard, $\log[h(t)/h_0(t)]$, which is often utilized for assessing calibration of the predictive models \citep{pi1, pi2, vanhoulingen}. A higher value indicates a worse prognosis when the event is an adverse outcome, such as death or relapse of disease. 
    \label{remark1}
    
    \item[2.] It is straightforward that the estimated log of relative hazard in \eqref{EG.cox3} becomes equivalent to $\bb{\widehat\beta}_U^{\T} \bb U + \bb{\widehat\beta}_V^\T \bb V$, since $\bb{\widehat\beta}_U = \widehat\theta_0 \cdot (\bb{\widehat\eta}, \bb{\widehat\gamma})$ and $\bb{\widehat\beta}_V = \bb{\widehat\theta}_1$.
   \label{remark2}

    \item[3.] In the absence of domain knowledge in the first stage, which precludes the decomposition of $\bb U$ into two parts, one may use the model $h(t) = h_0(t) \exp (\bb\gamma^\T \bb U)$ instead of \eqref{EG.cox}. In this case, the penalized Cox regression model is fitted by shrinking all coefficients associated with $\bb U$.
    \label{remark3}

    \item[4.] All $\bb V$ variables are included as covariates in \eqref{EG.cox3}, since $\bb V$ is usually low-dimensional and contains key variables. However, if a partial set of $\bb V$ is known to be clinically irrelevant, one can drop such variables from the model. When there is no prior knowledge about the clinical importance of the entire $\bb V$ variables, a simple remedy is to perform a partially penalized Cox regression, similar to the first stage, to retain the prognostic index $\widehat{\zeta}$ and assess $\bb V$ in a data-driven manner. In the extreme case where $\bb V$ is not only one-dimensional but also known to be clinically irrelevant, we suggest merely taking  $\widehat{\zeta}$ in the second stage. 
    \label{remark4}

\end{enumerate}

\section{Simulation experiments} 
\label{sec:simul}

In this section, simulation studies are conducted to compare the proposed approach to other existing methods and to evaluate the model performance using various metrics. 


\subsection{A binary missing covariate}
\label{binaryV}

\subsubsection{Simulation setup}
The survival data are generated from a Weibull distribution with shape parameter 1 and scale parameter 1, following the method of \citet{bender}{\color{black}, based on \eqref{regular.cox}.} The censoring times are simulated according to an exponential distribution with parameter $c_0$, which corresponds to a censoring rate of 80\%.
Traditional covariates $\bb U$ are $p$-dimensional standard normal random vector. A binary indicator $\bb V \in \{0,1\}$ is randomly generated with success probability conditional on the first variable of $\bb U$, which varies as follows: $\pi = 0.3$ if $U_1 < -0.5$; $\pi = 0.5$ if $U_1 \in [-0.5, 1)$; and $\pi = 0.2$ if $U_1 \geq 1$. 
{\color{black} Data-generating coefficients are set as $\bb\beta_V = 1.25$, along with three different $\bb\beta_U$ cases: $\bb\beta_U = (0.5_{p/2}, 0_{p/2})$ for weak dense (Scenario I);  $\bb\beta_U = (1.25, 1, 0.75, 0_{p-3})$ for strong signal (Scenario II); and $\bb\beta_U = (0.75, 0, 0.75_{.2p}, 0_{.8p-2})$ for moderate concentration (Scenario III)}. To create two-phase data with $\bb V$ being partially available such that $n'/n \approx 0.3$, we consider {\color{black} three settings} for the missing mechanism: (i) missing completely at random (MCAR); (ii) missing at random (MAR); {\color{black} and (iii) MAR with a mild-to-moderate violation. 
Details of these settings are provided in Web Appendix B.}


For comparison, we analyze the data using our proposed expert-guided (EG) method, complete-case analysis (CCA), na\"ive imputation (NI) with mode imputation, multiple imputation (MI) following \citet{wood2008} {\color{black} (hereafter referred to as MI-Wood), and improved MI by \citet{bartlett2015multiple} using rejection sampling (hereafter referred to as MI-Bartlett).
For the MI methods, we consider $K=5$, where a variable is considered selected if it appears in at least half of the five imputed datasets.
}
The penalty function that we consider across all methods is adaptive Lasso. 
The 5-fold cross-validation is used to select an optimal $\lambda_n$ for each penalization method. For the proposed EG method, we assume the existence of domain knowledge regarding the clinical importance of $\bb V$ and the first two variables in $\bb U$. The predictive model performance is mainly assessed using the c-index, calibration slope, and integrated Brier score (IBS). 
Variable selection performance on $(\bb\beta_U, \bb\beta_V)$ is evaluated by the Matthews correlation coefficient (MCC). {\color{black} A detailed description of each metric is provided in Web Appendix C.}
The performance measures are evaluated on an independent test set of size $n'$.
{\color{black} In Web Appendix D, we repeat the analysis by applying domain knowledge to the comparison methods as well.}

\subsubsection{Simulation results} 

\noindent 
Model performance metrics are presented with $n=150$ and $p=10$ based on 100 replications.
Under the MCAR setting (Table \ref{table:MCAR}), in almost all cases, 
our proposed EG method outperformed its counterparts in terms of a higher c-index, better calibration, lower IBS, and higher MCC. {\color{black} In contrast, while the NI and the two MI approaches generally performed better than the CCA in terms of c-index, their calibration slopes deviated from the ideal value of 1, signaling inaccurate risk estimates. The MI-Bartlett method outperformed the MI-Wood method in terms of c-index, IBS, and MCC, as missing values were imputed from models which are compatible with the substantive model using rejection sampling. However, the MI-Bartlett method had an issue of overfitting, as indicated by a calibration slope greater than 1 with increased variability. For example, in Scenario I, the MI-Bartlett method produced a calibration slope of 4.72, which deviates substantially from the ideal value of 1.
The superiority of the proposed EG method was most apparent in Scenario I, followed by a smaller, yet still noticeable, improvement in Scenario II. In Scenario III, the MI-Bartlett method showed a similar c-index to the EG method, with better IBS and/or MCC. This was expected, as the first two variables of $\bb U$ were included in the model based on domain knowledge, making the proposed method less suitable when $U_2$ had no effect. Nonetheless, the EG method consistently provided the best calibration, along with its lower standard deviation, which was nearly one-tenth that of the MI-Bartlett method. Calibration was consistently enhanced using the EG method across various scenarios by integrating the prognostic index and effectively utilizing two-phase data.}

\begin{table}[!ht]
  \begin{center}
  \caption{Simulation results under the MCAR setting. For each performance metric, the mean is reported with the standard deviation in parentheses. The best results are highlighted in boldface.}
\label{table:MCAR}
\begin{threeparttable}
  \fontsize{10}{11}\selectfont
  \begin{tabularx}{.76\textwidth}{ll| cccc}
    \toprule
    \multirow{2}{*}{Scenario} & \multirow{2}{*}{Method} &  \multirow{2}{*}{c-index}  &  Calibration  &  \multirow{2}{*}{IBS $\times 10^{1}$} & \multirow{2}{*}{MCC} \\
    & & & slope  &  & \\
    \midrule
    I   & CCA & 0.56 (0.11) & 1.29 (1.64) & 2.25 (0.58) & 0.10 (0.19) \\
& NI & 0.61 (0.13) & 1.98 (3.16) & 2.19 (0.62) & 0.25 (0.27) \\
& MI-Wood & 0.59 (0.11) & 2.39 (3.71) & 2.22 (0.56) & 0.20 (0.22) \\
& MI-Bartlett & 0.62 (0.12) & 4.72 (11.3) & 2.21 (0.60) & 0.31 (0.24) \\
& EG & \textbf{0.71} (0.11) & \textbf{0.94} (0.86) & \textbf{2.07} (0.67) & \textbf{0.62} (0.11) \\
    II  & CCA & 0.66 (0.15) & 1.61 (1.51) & 2.12 (0.80) & 0.34 (0.33) \\
& NI & 0.81 (0.09) & 1.91 (1.30) & 1.75 (0.76) & 0.72 (0.19) \\
& MI-Wood & 0.80 (0.09) & 2.30 (2.04) & 1.78 (0.75) & 0.68 (0.15) \\
& MI-Bartlett & 0.81 (0.09) & 2.24 (1.61) & 1.75 (0.78) & 0.74 (0.17) \\
& EG & \textbf{0.82} (0.09) & \textbf{0.96} (0.54) & \textbf{1.74} (0.98) & \textbf{0.85} (0.09) \\
    III & CCA & 0.60 (0.12) & 1.56 (1.29) & 2.27 (0.81) & 0.24 (0.32) \\
& NI & 0.68 (0.14) & 2.17 (2.15) & 2.07 (0.73) & 0.54 (0.33) \\
& MI-Wood & 0.67 (0.14) & 2.64 (3.86) & 2.05 (0.66) & 0.50 (0.29) \\
& MI-Bartlett & 0.71 (0.14) & 4.14 (7.40) & \textbf{1.99} (0.67) & \textbf{0.62} (0.28) \\
& EG & \textbf{0.72} (0.12) & \textbf{0.84} (0.65) & 2.01 (0.78) & 0.56 (0.18) \\
    \bottomrule
  \end{tabularx}
  \end{threeparttable}
  \end{center}
\end{table}

Under the setting of MAR (Table \ref{table:MAR}), the proposed EG method still outperformed the other alternatives in most cases, demonstrating a higher c-index, a calibration slope closer to 1, a smaller IBS, and a higher MCC. 
{\color{black}
Additionally, lower variability in the calibration slope was observed for the EG method. In contrast, the NI and the two MI approaches exhibited poor calibration, with an increased standard deviation observed particularly for both the MI-Wood and MI-Bartlett methods.}
Substantial variation in the final model with the MI approaches might have been caused by large fluctuations in variable selection results across multiply imputed datasets, in the coefficients of selected variables, or in both. This highlights a major challenge for MI methods when variable selection is involved.


\begin{table}[!ht]
  \begin{center}
  \caption{Simulation results under the MAR setting. For each performance metric, the mean is reported with the standard deviation in parentheses. The best results are highlighted in boldface.}
\label{table:MAR}
\begin{threeparttable}
  \fontsize{10}{11}\selectfont
  \begin{tabularx}{.76\textwidth}{ll| cccc}
    \toprule
    \multirow{2}{*}{Scenario} & \multirow{2}{*}{Method} &  \multirow{2}{*}{c-index}  &  Calibration  &  \multirow{2}{*}{IBS $\times 10^{1}$} & \multirow{2}{*}{MCC} \\
    & & & slope  &  & \\
    \midrule
    I   & CCA & 0.57 (0.11) & 2.27 (4.97) & 2.25 (0.77) & 0.09 (0.19) \\
& NI & 0.59 (0.13) & 2.06 (3.14) & 2.14 (0.69) & 0.23 (0.26) \\
& MI-Wood & 0.57 (0.11) & 1.93 (3.41) & 2.22 (0.66) & 0.18 (0.23) \\
& MI-Bartlett & 0.60 (0.13) & 5.51 (16.6) & 2.21 (0.73) & 0.29 (0.24) \\
& EG & \textbf{0.71} (0.12) & \textbf{0.88} (0.75) & \textbf{1.94} (0.84) & \textbf{0.63} (0.12) \\
    II  & CCA & 0.73 (0.15) & 2.42 (4.63) & 1.79 (0.78) & 0.44 (0.28) \\
& NI & 0.84 (0.08) & 3.05 (9.57) & 1.41 (0.69) & 0.70 (0.17) \\
& MI-Wood & 0.83 (0.09) & 3.30 (9.57) & 1.46 (0.71) & 0.67 (0.16) \\
& MI-Bartlett & 0.84 (0.08) & 3.20 (9.42) & 1.44 (0.71) & 0.73 (0.15) \\
& EG & \textbf{0.85} (0.08) & \textbf{1.05} (0.76) & \textbf{1.36} (0.64) & \textbf{0.85} (0.09) \\
    III & CCA & 0.60 (0.14) & \textbf{1.05} (1.69) & 2.12 (0.79) & 0.27 (0.28) \\
& NI & 0.69 (0.15) & 2.05 (1.78) & 1.86 (0.78) & 0.52 (0.34) \\
& MI-Wood & 0.68 (0.15) & 2.84 (3.31) & 1.85 (0.70) & 0.49 (0.31) \\
& MI-Bartlett & 0.71 (0.13) & 3.92 (9.29) & 1.84 (0.76) & \textbf{0.61} (0.30) \\
& EG & \textbf{0.73} (0.13) & 0.81 (0.70) & \textbf{1.81} (0.84) & 0.55 (0.18) \\
    \bottomrule
  \end{tabularx}
  \end{threeparttable}
  \end{center}
\end{table}

{\color{black}
Table \ref{table:MARviol} presents the performance metrics under a mild-to-moderate violation of the MAR assumption. The c-index for the CCA, NI, and the two MI methods remained similar or had a slight decrease compared to the MAR setting, while the calibration slope increased in nearly all cases, particularly for MI-Bartlett, which deviated further from 1. In contrast, the EG method maintained consistent performance, preserving its superior performance.
}


\begin{table}[!ht]
  \begin{center}
  \caption{Simulation results under the MAR setting with a mild-to-moderate violation. For each performance metric, the mean is reported with the standard deviation in parentheses. The best results are highlighted in boldface.}
\label{table:MARviol}
\begin{threeparttable}
  \fontsize{10}{11}\selectfont
  \begin{tabularx}{.76\textwidth}{ll| cccc}
    \toprule
    \multirow{2}{*}{Scenario} & \multirow{2}{*}{Method} &  \multirow{2}{*}{c-index}  &  Calibration  &  \multirow{2}{*}{IBS $\times 10^{1}$} & \multirow{2}{*}{MCC} \\
    & & & slope  &  & \\
    \midrule
    I   & CCA & 0.56 (0.10) & 3.66 (15.4) & 2.28 (0.76) & 0.11 (0.20) \\
& NI & 0.58 (0.12) & 1.42 (1.84) & 2.19 (0.66) & 0.23 (0.26) \\
& MI-Wood & 0.58 (0.11) & 2.31 (3.79) & 2.20 (0.65) & 0.19 (0.22) \\
& MI-Bartlett & 0.61 (0.13) & 6.97 (17.5) & 2.22 (0.87) & 0.30 (0.24) \\
& EG & \textbf{0.71} (0.12) & \textbf{0.91} (0.88) & \textbf{1.93} (0.80) & \textbf{0.63} (0.12) \\
    II  & CCA & 0.72 (0.15) & 2.04 (3.39) & 1.83 (0.78) & 0.40 (0.29) \\
& NI & 0.84 (0.09) & 3.58 (13.9) & 1.43 (0.69) & 0.70 (0.17) \\
& MI-Wood & 0.83 (0.09) & 3.14 (9.24) & 1.48 (0.72) & 0.68 (0.13) \\
& MI-Bartlett & 0.83 (0.09) & 3.37 (11.3) & 1.47 (0.71) & 0.73 (0.17) \\
& EG & \textbf{0.85} (0.08) & \textbf{1.06} (0.77) & \textbf{1.40} (0.66) & \textbf{0.85} (0.09) \\
    III & CCA & 0.62 (0.14) & 1.65 (2.96) & 2.13 (0.90) & 0.28 (0.28) \\
& NI & 0.69 (0.14) & 2.07 (1.83) & 1.87 (0.76) & 0.51 (0.33) \\
& MI-Wood & 0.69 (0.15) & 2.79 (2.87) & 1.87 (0.74) & 0.50 (0.30) \\
& MI-Bartlett & 0.72 (0.14) & 4.24 (10.1) & \textbf{1.85} (0.76) & \textbf{0.60} (0.27) \\
& EG & \textbf{0.73} (0.14) & \textbf{0.75} (0.67) & 1.90 (0.95) & 0.56 (0.18) \\
    \bottomrule
  \end{tabularx}
  \end{threeparttable}
  \end{center}
\end{table}

{\color{black}
}

\subsubsection{Impact of the sample size ratio in two-phase data}
\label{sec:impact}
\noindent
{\color{black}
We conduct additional simulations to further illustrate the impact of the sample size ratio in two-phase data, i.e., $n'/n$. Due to the characteristics of two-phase data (see Figure \ref{fig:basic}), $n'$ is expected to be low, as it typically involves expensive biomarkers. To investigate this, we increase the sample size to $n=1,000$, allowing $n'$ to vary within a reasonable range that ensures both sufficient number of samples and number of events, 
where $n'/n \approx \{0.10, 0.15, 0.25\} = r$ (see Web Appendix E).
When $r = 0.10$ (Web Table 4), our EG method performed the best. The benefit of our proposed method as compared to the method that discards individuals with missing information (e.g., CCA) was especially evident when the number of the target samples is substantially limited. As $r$ increased to $0.15$ (Web Table 5) and $0.25$ (Web Table 6), the difference between our method and the competing methods gradually decreased.
However, our method consistently outperformed the alternative approaches in terms of c-index, calibration slope, IBS, and MCC. Additionally, it had a calibration slope very close to 1, regardless of the value of $r$. Thus, the prognostic index contributes to well-calibrated risk predictions, a feature that none of the competing methods achieved.


}

\subsection{A continuous missing covariate}
\label{continuousV}

{\color{black} 
Additional simulations are conducted to evaluate our proposed method in comparison to other methods with a continuous missing covariate. The data-generating distribution and setup resembles Section \ref{binaryV} except for $\bb V$, where $\bb V \sim N(0.4|U_1| - 0.1, 0.2^2)$, and the types of models to be used for imputing variables, including mean imputation for NI and the default imputation methods for MI approaches. Under this simulation design (Web Appendix F), the results remained largely the same as in the binary missing covariate case, demonstrating that our proposed method outperformed its alternatives in terms of a higher c-index, calibration slope closer to 1, lower IBS, and higher MCC in a majority of cases (Web Tables 7--9).
}




\section{Real data application}
\label{sec:real}

In this section, we apply our proposed model to the data from \citet{wotman2019}, which comprises patients diagnosed with NPC between 2013--2015 with 3 years of follow-up extracted from the SEER database.
Histologically confirmed NPC patients were included using the International Classification of Diseases for Oncology, Third Edition (ICD-O-3), with topography codes for histologic types including 8070, 8071, 8072, 8073, 8020, 8021, and 8082, and site of origin codes including 110, 111, 112, 113, 118, and 119. Patients having malignancies of other head and neck subsites and those with missing follow-up information were excluded from the analysis. Thus, a total of 1,762 NPC patients were included in this analysis with a median follow-up time of 11 months, in which 266 of them died due to NPC. 
There were 1,245 NPC patients whose HPV status was unknown, leaving only 517 with known HPV status, of whom 180 were tested positive for HPV.
Outcome of interest was cause-specific survival in months, which is time from cancer diagnosis to death due to NPC. 
Patients who died of a cause other than NPC or who were alive by the end of study or lost to follow-up were considered censored at the last date on which they were known to be alive.

\begin{figure}
\begin{subfigure}{.5\linewidth}
\centering
\includegraphics[scale=.31]{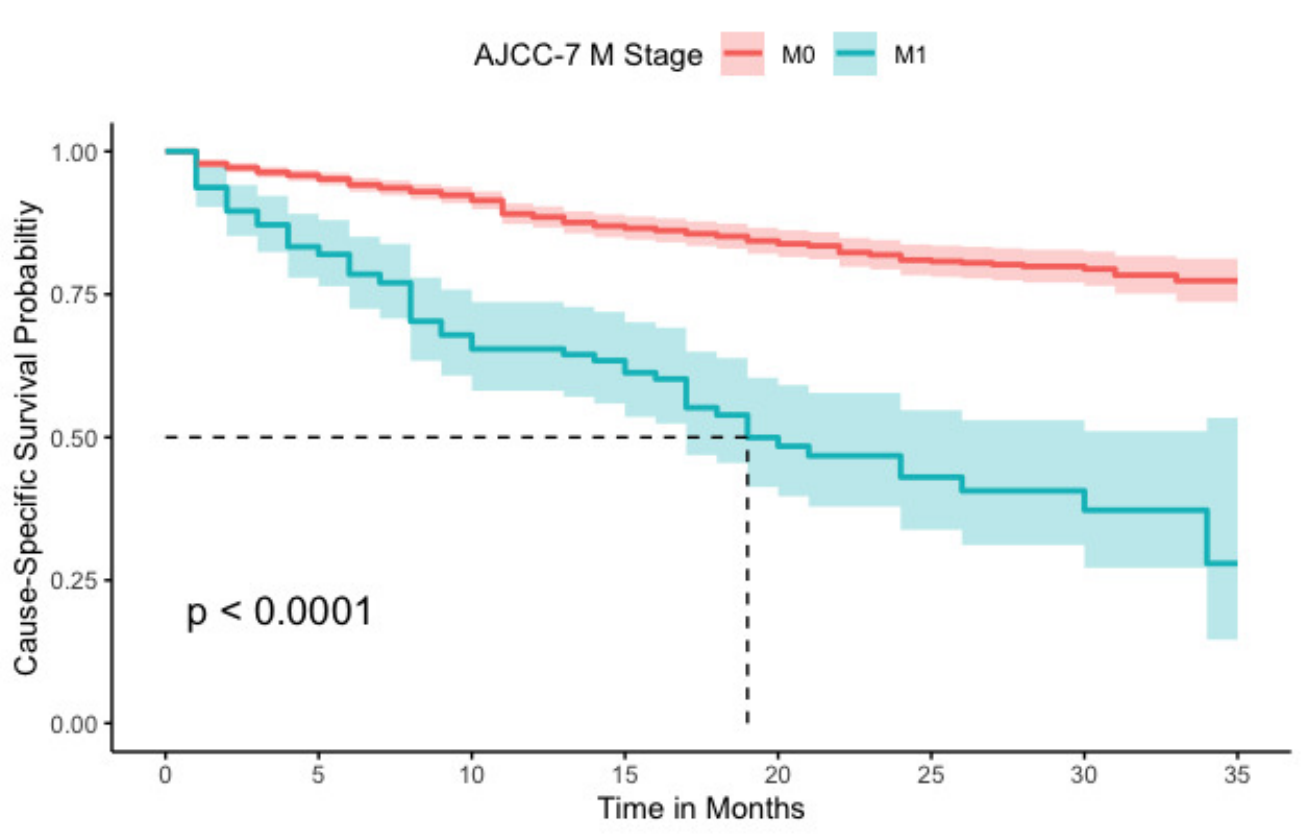}
\caption{}
\label{fig:sub1}
\end{subfigure}%
\begin{subfigure}{.5\linewidth}
\centering
\includegraphics[scale=.31]{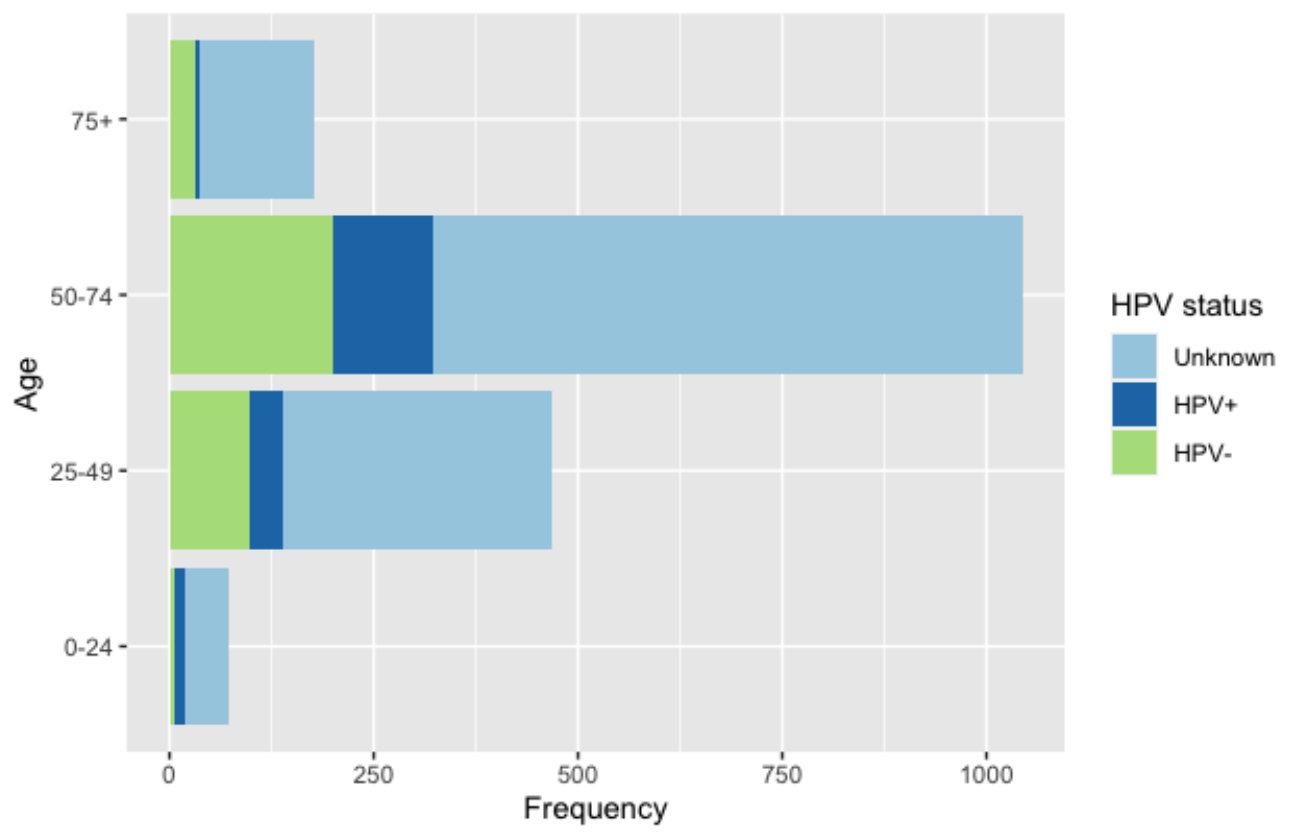}
\caption{}
\label{fig:sub2}
\end{subfigure}\\[1ex]
\begin{subfigure}{\linewidth}
\centering
\includegraphics[scale=.45]{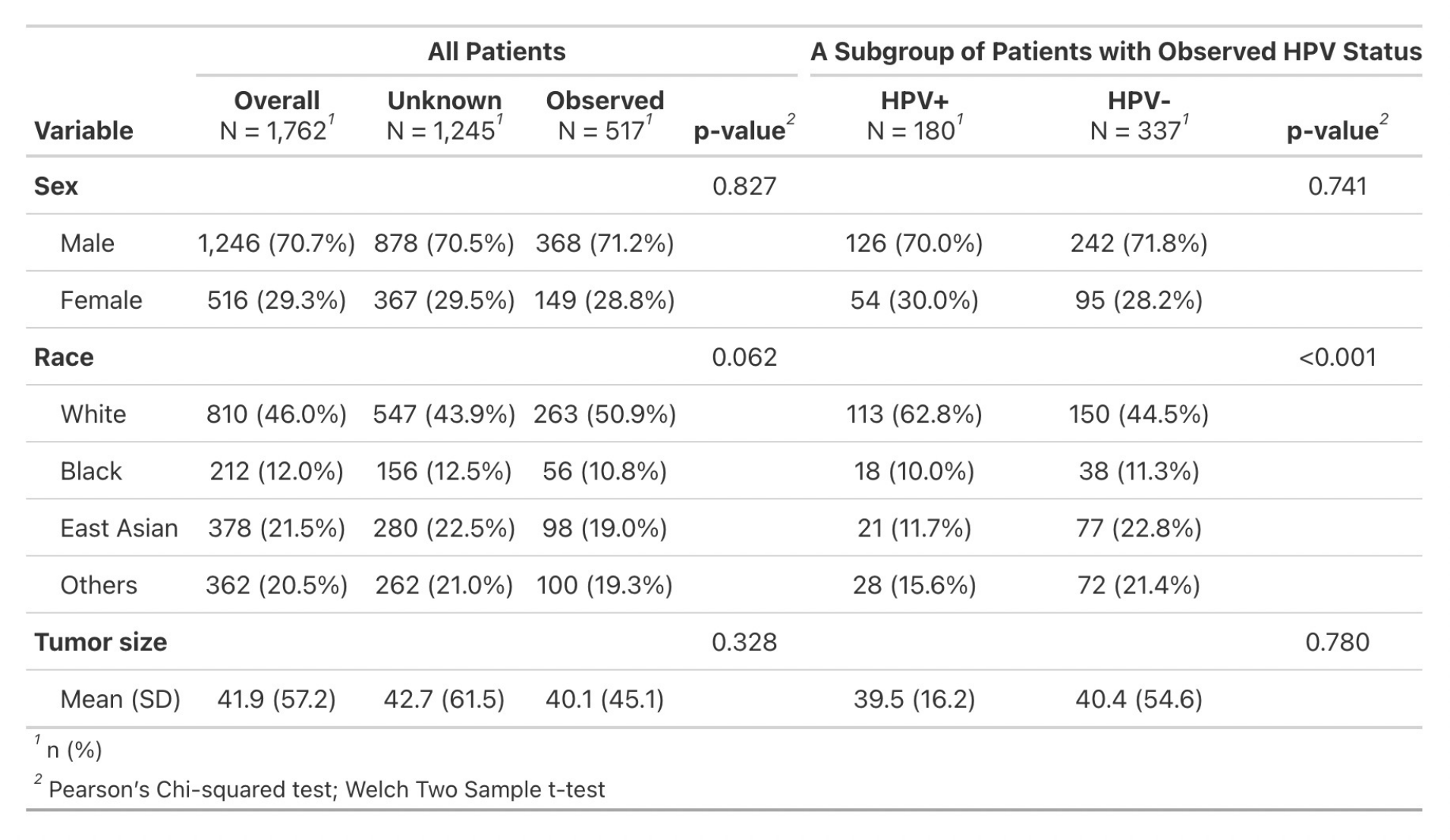}
\caption{}
\label{fig:sub3}
\end{subfigure}
\caption{Descriptive analyses of the study cohort: (A) Kaplan-Meier curves by AJCC-7 M stage for cause-specific survival; (B) age distribution by HPV status; (C) a summary statistics in terms of age, race, and tumor size for all patients (stratified by HPV status unknown vs. observed) and for a subgroup of patients with observed HPV only (stratified by HPV+ vs. HPV--).
}
\label{fig:NPC}
\end{figure}

Traditional baseline covariates $\bb U$ included gender (male or female), age ($<$25, 25--39, 40--54, 55--69, 70--84, or 85+), marital status (married, single, or other), race (White, Black, East Asians, or other), histology (keratinizing or non-keratinizing), AJCC-7 stage (I/II or III/IV), AJCC-7 T stage (early or advanced), AJCC-7 N stage (0 or 1+), AJCC-7 M stage (M0 or M1), sequence of primary disease (one primary only or other), and tumor size. The HPV status with nearly a 70\% missing rate is denoted as $\bb V$ in this example.
Figure \ref{fig:NPC} presents clinical and demographic characteristics of the study cohort in terms of age, race, sex, HPV status, AJCC-7 M stage, and tumor size. Figure \ref{fig:sub1} depicts a na\"ive comparison of cause-specific survival between two groups of AJCC-7 M stage. Figure \ref{fig:sub2} demonstrates that age distribution by HPV status. Figure \ref{fig:sub3} characterizes differences in sex, race, and tumor size using all patients stratified by whether HPV status was observed or not and using a subgroup of patients with observed HPV stratified by whether a patient is HPV+ or HPV--.
A full descriptive statistics of the study samples are provided in Web Appendix I. 

The proposed method was applied and compared with other methods as in Section \ref{sec:simul}. 
The c-index, calibration slope, and IBS were assessed to compare the predictive ability of different models for cause-specific survival in NPC patients. We used 5-fold cross-validation on the target samples of size $n'$ to create a hold-out test set (20\%) for each fold. The remaining 80\% of $n'$ samples, along with $n-n'$ samples that had missing HPV status, constituted the training set. We developed a model on each training set and evaluated the performance metrics on each test set. The performance measures were then averaged to obtain an overall estimate of the model performance.  
For our method, we penalized all $\bb U$ variables except for AJCC-7 stage in the first stage based on discussions with domain experts in nasopharyngeal oncology. Additionally, since the existing studies \citep{verma2018, wotman2019,  jiang2016, wu2021} have linked the prognostic role of HPV to NPC, 
the HPV status, denoted as $\bb V$, was included in the second stage along with the prognostic index. 

The results are summarized in Table \ref{table:npc}. Notably, our proposed EG method achieved a higher c-index, a calibration slope closer to 1, and a lower IBS compared to the other methods. The largest difference was observed between the CCA and EG methods, highlighting the disadvantage of analyzing patients with known HPV status only, which incurs a substantial loss of information. {\color{black} The NI the MI methods achieved a c-index of 0.76 and 0.77, respectively. In contrast, the CCA had the lowest c-index of 0.53, while the EG method outperformed all others with the highest c-index of 0.81. The CCA and all three imputation methods did not produce well-calibrated risk estimates, as indicated by the calibration slope deviated from the target value of 1. In particular, the unbiasedness of risk estimation was notable with the proposed EG method which had a calibration slope of 1.07.}

\begin{table}[!ht]
  \centering
  \caption{Performance results of methods applied to NPC data. The best results are highlighted in boldface.}
  \begin{threeparttable}
  \fontsize{10}{11}\selectfont
  \begin{tabular}{ l >{\centering\arraybackslash}p{1.8cm} >{\centering\arraybackslash}p{1.8cm} >{\centering\arraybackslash}p{1.8cm} >{\centering\arraybackslash}p{1.8cm} >{\centering\arraybackslash}p{1.8cm} }
  \toprule
          & CCA & NI & MI-Wood & MI-Bartlett & EG \\
  \midrule
  c-index & 0.53 & 0.76 & 0.77 & 0.77 & \textbf{0.81} \\
  Calibration slope \qquad\qquad\qquad & 2.67 & 1.14 & 1.20 & 1.20 & \textbf{1.07} \\
  IBS $\times 10^1$ & 1.30 & 1.23 & 1.21 & 1.21 & \textbf{1.14} \\
\bottomrule
   \end{tabular}
    \end{threeparttable}
    \label{table:npc}
\end{table}

The proposed EG method, which has shown good discrimination and reliable risk estimates, can serve as a key to successful risk stratification and risk-guided clinical decision-making for NPC patients. To demonstrate its utility for risk stratification, we trained our proposed model using the entire dataset and finally applied it to the target samples to divide them into low-, medium-, and high-risk groups. 
As shown in Web Figure 1, there was a significant difference in cause-specific survival probabilities across these groups ($p < .001$) based on the log-rank test. The estimated 2-year cause-specific survival (95\% CI) was 93.3\% (85.8\%, 100.0\%) for the low-risk group, 85.1\% (79.0\%, 91.7\%) for the medium-risk group, and 58.7\% (46.9\%, 73.5\%) for the high-risk group. 
A pairwise log-rank test with the Bonferroni-Holm method of adjustment indicated that there were significant pairwise differences between all three groups. 
Our study results will help guide therapeutic strategies in clinical practice, including radiation therapy, chemotherapy, or a combination of both, as well as different follow-up care options (e.g., intensive vs. regular monitoring) across various risk groups.

\section{Discussion}
\label{sec:discuss}

Accurate survival predictions based on a stable and efficient model enable cancer patients to proactively make plans for their death and achieve goal-concordant care. However, limited efforts have been made towards model updating or leveraging information especially for the two-phase data with survival outcomes. In this paper, we proposed the expert-guided penalized Cox regression model to address several important issues pertinent to the NPC data. Using this method, we efficiently leveraged two-phase data and incorporated domain knowledge into a data-driven variable selection procedure which led to improved prediction of survival outcomes. 
Additionally, the re-calibration process helped eliminate the over- or under-estimation of risk in the target samples. 
{\color{black} 
In contrast, the MI-Bartlett method had overfitting issues, demonstrated by a calibration slope substantially greater than 1, despite having better discriminatory ability compared to the MI-Wood method. While increasing the complexity of the imputation method may improve the model, it also introduces the risk of overfitting when applied to new datasets. It is important to note that when the model fits the training data extremely well with increased complexity, its ability to perform effectively on the independent test data may be compromised.}

Through a series of numerical studies and real data application, we have observed that our proposed method achieve a higher c-index, a calibration slope in close proximity of the ideal value of 1, and a lower IBS. {\color{black} 
To further demonstrate the utility of our approach, we conducted additional simulations with two missing covariates (Web Appendix G). The results largely remained consistent, demonstrating that our proposed method outperformed its alternatives.
Additional simulations are warranted to gain a comprehensive understanding of the impact of the missing covariate dimension in two-phase data, by further increasing the number of missing covariates. However, continuously expanding the dimension of $\bb V$ becomes less practical in real-world clinical settings, as the resources required to obtain expensive biomarkers may often be limited, not only by the number of individuals that are available for the study, but also by the number of variables that can feasibly be retrieved.}

While the prognostic index is re-calibrated along with key factors to adapt to the target samples, a more refined version of the proposed method could allow the prognostic index to differ between the target samples and the remaining ones in the initial step. It is also crucial to recognize that the proposed method can accommodate a variety of penalty functions, although we have primarily focused on adaptive Lasso due to its oracle property. Furthermore, in the absence of domain knowledge about the clinical importance of some variables, the proposed model can still be applied with a slight modification, as illustrated in the Remarks in Section \ref{subsec:method}. The flexibility is one of the major advantages of the proposal. 

The proposed method with more reliable and accurate risk estimates can lead to the successful risk-stratified care and clinical support. Especially for cancer patients at the terminal stage, accurate survival estimation will help prevent the overuse of aggressive treatments and reduce unnecessary toxicity. 
{\color{black}
Improved survival estimation is guaranteed only when the proportional hazards assumption holds, which we have assumed throughout the paper. To evaluate the impact of non-proportionality of hazards, additional simulations were conducted (see Web Appendix H). The results showed a reduced c-index and biased risk estimates across all methods, as expected, although the overall impact on our method was relatively small. The variable selection performance was significantly impacted for all methods except our method. Our proposed method experienced only a minimal decrease in MCC, due to the benefit of incorporating domain knowledge and prognostic index.
Future research could explore more flexible models that do not rely on the proportional hazards assumption.} 
More complex missing data patterns would further complicate model building and evaluation. We plan to pursue extensions of our proposed work along these lines in future research.



\section*{Acknowledgements}

We thank the reviewers, the associate editor, and the co-editor for their careful review and thoughtful feedback.

\section*{Funding}

None declared.

\section*{Conflict of interest}

None declared.

\section*{Data availability}

The data that supports the findings in this paper were obtained from the Surveillance, Epidemiology, and End Results (SEER) database of the National Cancer Institute. The SEER data are publicly available at https://seer.cancer.gov. Interested researchers can request access by signing a SEER Research Data Agreement. 




\section*{Supplementary data}

Web Appendices, Tables, and Figures referenced in Sections \ref{sec:simul}--\ref{sec:discuss} and R codes are available with this paper at the Biometrics website on Oxford Academic. The codes are also available on Github: https://github.com/oheunj/TwoPhaseSurv.

\bibliographystyle{biom} \bibliography{references}

\end{document}





\if0\blind
{
  \bigskip
  \bigskip
  \bigskip
  \begin{center}
    {\LARGE Supplementary Materials for ``Leveraging two-phase data for improved prediction of survival outcomes with application to nasopharyngeal cancer'' by Eun Jeong Oh, Seungjun Ahn, Tristan Tham, and Min Qian}
\end{center}
  \medskip
} \fi

\renewcommand{\thesection}{Web Appendix \Alph{section}}


\setcounter{table}{0} 

\captionsetup[table]{name=Web Table,          
                     skip=1ex, labelfont=bf}

\renewcommand{\figurename}{Web Figure}

\setcounter{figure}{0} 



\bigskip\bigskip\bigskip

\section{}

The fact that a small subset of patients are being tested for expensive biomarkers may suggest that there is a systematic reason for missing data on those biomarkers. Possible explanations could be that only a subset of patients are eligible or willing to participate, or certain characteristics (e.g., age, cancer stage) may increase the likelihood of undergoing the test. If the missingness is related to these observed characteristics but not to the missing values themselves, the missing data would be  classified as Missing at Random (MAR). For instance, in the NPC study, patients nearing the end of their lives at terminal stages typically do not undergo HPV testing, as such tests would not improve their condition. Therefore, it is reasonable to assume MAR, as the missingness is related to an observable characteristic (cancer stage). Missing Not at Random (MNAR) may occur when the missing data is related to the missing values themselves. However, identifying MNAR is challenging because it depends on unobserved information, making it difficult to infer without deep insights or auxiliary data.

As part of a cost-effective two-phase sampling design, if a sub-sample of patients is randomly selected to undergo additional testing, this would represent Missing Completely at Random (MCAR). In this case, the missing data on the additional variables is assumed to be unrelated to both observed factors and the values that would have been observed had the data not been missing, as the selection for further testing is purely random and not influenced by any specific characteristics of the patients. Conversely, if the selection is impacted by observed data but not by the missing values themselves, the missing data would be classified as MAR. An example of this occurs when patients with insurance coverage are more likely to undergo further biomarker testing, such that the missingness of biomarker data is dependent on the insurance status (an observed variable) but not on the missing biomarker values themselves.

\section{}

Here we describe the complete process for generating two phase data under different missing mechanisms as described in Section 3.1. We consider three settings for the missing mechanism: (i) missing completely at random (MCAR) where the probability of $\bb V$ being missing is independent of all covariates; (ii) missing at random (MAR) where the probability of $\bb V$ being missing depends on the observed data; and (iii) MAR with a mild-to-moderate violation, where the probability of $\bb V$ being missing remains dependent on the observed data, but is slightly modified based on the missing data.
For each desired ratio $r \approx n'/n$, the probability of $\bb V$ being missing is $1-r$ for (i) MCAR, $(1-r)/\Phi (-\Phi^{-1} (r/3))$ if $U_1 > \Phi^{-1}(r/3)$ and $0$ otherwise for (ii) MAR, and $(1- r - 0.1 I(\bb V \leq 0))/\Phi (-\Phi^{-1} (r/3)) + 0.1 I(\bb V > 0)$ if $U_1 > \Phi^{-1}(r/3)$ and 0.1 otherwise for (iii) MAR with a mild-to-moderate violation, where $\Phi(\cdot)$ is the cumulative distribution function of the standard normal distribution. 
This ensures approximately $n'=r \cdot n$ across all three settings.

\section{}

Hereby we differentiate discrimination, calibration, and overall performance of a survival model. 

\begin{itemize}
    \item[] \textit{Discrimination.} A model should be able to accurately discriminate different risk categories. Discrimination indicates how well a model can distinguish between patients who will die earlier and those who would die later. The Harrell's c-index \citep{harrell1, harrell2, harrell3}, also known as concordance index or c-statistic, is a commonly used measure to evaluate a discriminate ability of survival models. The closer the c-index is to 1, the better the model discriminates between low-risk and high-risk patients. A value of 0.5 indicates that the model is no better out predicting outcomes than random chance. 
    \item[] \textit{Calibration.} The calibration of a model is a measure of an agreement between the observed and predicted outcomes. The commonly used calibration metric is calibration slope proposed by \citet{vanhoulingen}. A model that calibrates well would result in a calibration slope close to value 1. An overfitted model would have a slope $>$1, whereas an underfitted model would have a slope $<$1. Overfitting is more frequently observed, while underfitting occurs when a model is excessively simple. 
    \item[] \textit{Overall performance.} With regards to measuring overall performance, the commonly used metric is the Brier score proposed by \citet{brier1950}. It incorporates both discrimination and calibration aspects of a model, taking values between 0 and 1. The Brier score is similar to the mean squared error in linear regressions. The integrated Brier score (IBS), introduced by \citet{graf}, integrates multiple scores obtained at all follow-up times. A score closer to 0 implies a better predictive performance.
\end{itemize}

The predictive model performance is mainly assessed using the c-index, calibration slope, and IBS. 
Variable selection performance on $(\bb\beta_U, \bb\beta_V)$ is evaluated by the Matthews correlation coefficient (MCC) proposed by \citet{mcc}, defined as
\begin{align*}
    \text{MCC} = \frac{\text{TP} \cdot \text{TN} - \text{FP} \cdot \text{FN} }{\sqrt{(\text{TP}+\text{FP})\cdot(\text{TP}+\text{FN})\cdot(\text{TN}+\text{FP})\cdot(\text{TN}+\text{FN})}},
\end{align*}
where TP, TN, FP, and FN are true negatives, true negatives, false negatives, and false positives, respectively.

\section{}

In this section, we repeat the analysis in Section 3.1 by applying domain knowledge to the comparison methods as well. When the comparison methods were also applied with domain knowledge in the variable selection process, they demonstrated some benefits in general, as indicated by an increased c-index, lower IBS, improved calibration, and higher MCC (Web Tables \ref{table:MCAR}--\ref{table:MARviol}). However, despite noticeable improvements in calibration for the competing methods, the standard deviation of the calibration slope by the EG method was relatively lower than that of the other methods, suggesting that our method provided a more consistent estimate approaching the ideal value of 1. Overall, our proposed EG method still outperformed the alternatives in most cases.

Additionally, it is worth noting that the MCC for the comparison methods decreased in Scenario III. This was expected, as the first two variables of $\bb U$ were included in the model based on domain knowledge, which is suboptimal when $U_2$ had no effect. Thus, in this context, where all methods incorporated partly misaligned or incorrect domain knowledge, our proposed method, which initially had a lower MCC, turned out perform better in terms of variable selection compared to the other methods. Furthermore, it maintained good and consistent performance across various missing mechanisms in terms of c-index, calibration, and IBS, highlighting the robustness and reliability of the proposed method. \\

\begin{table}[!ht]
  \begin{center}
\caption{Simulation results under the MCAR setting when domain knowledge is also applied to the comparison methods. For each performance metric, the mean is reported with the standard deviation in parentheses. The best results are highlighted in boldface.}
\label{table:MCAR}
\begin{threeparttable}
  \fontsize{10}{11}\selectfont
  \begin{tabularx}{.76\textwidth}{ll| cccc}
    \toprule
    \multirow{2}{*}{Scenario} & \multirow{2}{*}{Method} &  \multirow{2}{*}{c-index}  &  Calibration  &  \multirow{2}{*}{IBS $\times 10^{1}$} & \multirow{2}{*}{MCC} \\
    & & & slope  &  & \\
    \midrule
    I   & CCA & 0.69 (0.11) & 0.81 (0.72) & 2.19 (0.71) & 0.58 (0.07) \\
& NI & 0.69 (0.12) & 1.53 (3.60) & 2.08 (0.61) & 0.50 (0.12) \\ 
& MI-Wood & 0.67 (0.11) & 1.50 (3.61) & 2.13 (0.57) & 0.47 (0.07) \\ 
& MI-Bartlett & 0.68 (0.11) & 1.26 (1.39) & 2.13 (0.58) & 0.51 (0.14) \\
& EG & \textbf{0.71} (0.11) & \textbf{0.94} (0.86) & \textbf{2.07} (0.67) & \textbf{0.62} (0.11) \\
    II  & CCA & 0.80 (0.10) & 0.83 (0.52) & 1.85 (1.01) & 0.82 (0.07) \\
& NI & \textbf{0.82} (0.08) & 1.15 (0.44) & \textbf{1.69} (0.76) & 0.71 (0.13) \\
& MI-Wood & 0.81 (0.08) & 1.14 (0.47) & 1.71 (0.77) & 0.67 (0.09) \\
& MI-Bartlett & 0.81 (0.08) & 1.15 (0.47) & 1.71 (0.77) & 0.69 (0.11) \\
& EG & \textbf{0.82} (0.09) & \textbf{0.96} (0.54) & 1.74 (0.98) & \textbf{0.85} (0.09) \\
    III & CCA & 0.69 (0.11) & 0.67 (0.71) & 2.14 (0.82) & 0.43 (0.12) \\
& NI & 0.70 (0.12) & 1.18 (1.02) & \textbf{1.99} (0.62) & 0.38 (0.24) \\
& MI-Wood & 0.68 (0.12) & \textbf{1.10} (0.94) & 2.04 (0.57) & 0.34 (0.21) \\
& MI-Bartlett & 0.70 (0.11) & 1.26 (0.93) & 2.00 (0.60) & 0.45 (0.22) \\
& EG & \textbf{0.72} (0.12) & 0.84 (0.65) & 2.01 (0.78) & \textbf{0.56} (0.18) \\
    \bottomrule
  \end{tabularx}
  \end{threeparttable}
  \end{center}
\end{table}

\begin{table}[!ht]
  \begin{center}
  \caption{Simulation results under the MAR setting when domain knowledge is also applied to the comparison methods. For each performance metric, the mean is reported with the standard deviation in parentheses. The best results are highlighted in boldface.}
\label{table:MAR}
\begin{threeparttable}
  \fontsize{10}{11}\selectfont
  \begin{tabularx}{.76\textwidth}{ll| cccc}
    \toprule
    \multirow{2}{*}{Scenario} & \multirow{2}{*}{Method} &  \multirow{2}{*}{c-index}  &  Calibration  &  \multirow{2}{*}{IBS $\times 10^{1}$} & \multirow{2}{*}{MCC} \\
    & & & slope  &  & \\
    \midrule
    I   & CCA & 0.70 (0.11) & 0.77 (0.55) & 2.01 (0.90) & 0.57 (0.06) \\
& NI & 0.68 (0.12) & 1.46 (3.77) & 1.99 (0.69) & 0.51 (0.13) \\
& MI-Wood & 0.67 (0.12) & 1.47 (3.76) & 2.00 (0.68) & 0.48 (0.10) \\
& MI-Bartlett & 0.67 (0.12) & 1.14 (1.03) & 2.00 (0.68) & 0.52 (0.14) \\
& EG & \textbf{0.71} (0.12) & \textbf{0.88} (0.75) & \textbf{1.94} (0.84) & \textbf{0.63} (0.12) \\
    II  & CCA & 0.84 (0.08) & 1.14 (2.23) & 1.45 (0.69) & 0.82 (0.07) \\
& NI & \textbf{0.85} (0.07) & 1.22 (0.63) & 1.40 (0.63) & 0.69 (0.13) \\
& MI-Wood & 0.84 (0.08) & 1.26 (0.87) & 1.43 (0.68) & 0.67 (0.08) \\
& MI-Bartlett & 0.84 (0.08) & 1.27 (0.88) & 1.42 (0.68) & 0.68 (0.11) \\
& EG & \textbf{0.85} (0.08) & \textbf{1.05} (0.76) & \textbf{1.36} (0.64) & \textbf{0.85} (0.09) \\
    III & CCA & 0.71 (0.13) & 0.68 (0.68) & 1.94 (0.88) & 0.41 (0.10) \\
& NI & 0.72 (0.13) & 1.20 (1.13) & \textbf{1.78} (0.72) & 0.39 (0.23) \\
& MI-Wood & 0.71 (0.14) & 1.21 (1.13) & 1.83 (0.71) & 0.34 (0.20) \\
& MI-Bartlett & 0.72 (0.13) & 1.27 (1.16) & 1.78 (0.70) & 0.44 (0.23) \\
& EG & \textbf{0.73} (0.13) & \textbf{0.81} (0.70) & 1.81 (0.84) & \textbf{0.55} (0.18) \\
    \bottomrule
  \end{tabularx}
  \end{threeparttable}
  \end{center}
\end{table}

\begin{table}[!ht]
  \begin{center}
  \caption{Simulation results under the MAR setting with a mild-to-moderate violation when domain knowledge is also applied to the comparison methods. For each performance metric, the mean is reported with the standard deviation in parentheses. The best results are highlighted in boldface.}
\label{table:MARviol}
\begin{threeparttable}
  \fontsize{10}{11}\selectfont
  \begin{tabularx}{.76\textwidth}{ll| cccc}
    \toprule
    \multirow{2}{*}{Scenario} & \multirow{2}{*}{Method} &  \multirow{2}{*}{c-index}  &  Calibration  &  \multirow{2}{*}{IBS $\times 10^{1}$} & \multirow{2}{*}{MCC} \\
    & & & slope  &  & \\
    \midrule
    I   & CCA & 0.70 (0.12) & 0.79 (0.71) & 2.04 (0.90) & 0.56 (0.05) \\
& NI & 0.68 (0.12) & 1.43 (3.54) & 1.99 (0.66) & 0.49 (0.12) \\
& MI-Wood & 0.67 (0.12) & 1.42 (3.55) & 1.99 (0.64) & 0.47 (0.09) \\
& MI-Bartlett & 0.67 (0.12) & 1.50 (3.55) & 1.97 (0.63) & 0.52 (0.13) \\
& EG & \textbf{0.71} (0.12) & \textbf{0.91} (0.88) & \textbf{1.93} (0.80) & \textbf{0.63} (0.12) \\
    II  & CCA & 0.84 (0.08) & 0.96 (0.64) & 1.49 (0.70) & 0.82 (0.06) \\
& NI & \textbf{0.85} (0.07) & 1.27 (0.79) & 1.43 (0.67) & 0.68 (0.12) \\
& MI-Wood & 0.84 (0.08) & 1.29 (0.85) & 1.44 (0.68) & 0.67 (0.09) \\
& MI-Bartlett & 0.84 (0.08) & 1.25 (0.66) & 1.44 (0.68) & 0.69 (0.11) \\
& EG & \textbf{0.85} (0.08) & \textbf{1.06} (0.77) & \textbf{1.40} (0.66) & \textbf{0.85} (0.09) \\
    III & CCA & 0.71 (0.13) & 4.79 (41.3) & 2.05 (1.00) & 0.40 (0.11) \\
& NI & 0.72 (0.13) & \textbf{1.18} (1.12) & 1.87 (0.75) & 0.38 (0.22) \\
& MI-Wood & 0.71 (0.13) & 1.23 (1.08) & 1.90 (0.74) & 0.33 (0.21) \\
& MI-Bartlett & 0.71 (0.13) & 1.23 (1.13) & 1.86 (0.73) & 0.44 (0.22) \\
& EG & \textbf{0.73} (0.14) & 0.75 (0.67) & \textbf{1.90} (0.95) & \textbf{0.56} (0.18) \\
    \bottomrule
  \end{tabularx}
  \end{threeparttable}
  \end{center}
\end{table}

\section{}

To understand the impact of the sample size ratio in two-phase data, we consider additional simulations to further illustrate the impact of the sample size ratio in two-phase data, i.e., $n'/n$. Due to the characteristics of two-phase data, $n'$ is expected to be low, as it typically involves expensive biomarkers. To investigate this, we increase the sample size to $n=1,000$, allowing $n'$ to vary within a reasonable range that ensures both sufficient number of samples and number of events in each fold of cross-validation,
where $n'/n \approx \{0.10, 0.15, 0.25\} = r$. 
When $r = 0.10$, our EG method still performed the best (Web Table \ref{table:r0.10}). The benefit of our proposed method as compared to the method that discards individuals with missing information (e.g., CCA) could be especially evident when the number of the target samples is substantially limited. As $r$ increased to $0.15$ (Web Table \ref{table:r0.15}) and $0.25$ (Web Table \ref{table:r0.25}), the difference between our method and the competing methods gradually decreased.
However, our method consistently outperformed the alternative approaches in terms of c-index, calibration slope, IBS, and MCC. Additionally, it had a calibration slope very close to 1, regardless of the value of $r$. Thus, the prognostic index contributes to well-calibrated risk predictions, a feature that none of the competing methods achieved.

\begin{table}[!ht]
  \begin{center}
  \caption{Simulation results for $r=0.10$ under various settings for missing mechanism. For each performance metric, the mean is reported with the standard deviation in parentheses. The best results are highlighted in boldface.}
\label{table:r0.10}
\begin{threeparttable}
  \fontsize{10}{11}\selectfont
  \begin{tabularx}{.87\textwidth}{lll| cccc}
    \toprule
    \multirow{2}{*}{Setting} & \multirow{2}{*}{Scenario}  &  \multirow{2}{*}{Method} &  \multirow{2}{*}{c-index}  &  Calibration  &  \multirow{2}{*}{IBS $\times 10^{1}$} & \multirow{2}{*}{MCC} \\
    & & & & slope  &  & \\
    \midrule
    MCAR & I   & CCA & 0.59 (0.10) & 1.49 (1.29) & 2.11 (0.43) & 0.24 (0.26)\\
        &    & NI & 0.75 (0.06) & 1.63 (0.57) & 1.76 (0.50) & 0.86 (0.10)\\
        &    & MI-Wood & 0.74 (0.06) & 1.65 (0.58) & 1.81 (0.45) & 0.81 (0.05)\\
        &    & MI-Bartlett & 0.74 (0.07) & 1.67 (0.60) & 1.81 (0.45) & 0.83 (0.07)\\
        &    & EG & \textbf{0.77} (0.07) & \textbf{0.93} (0.36) & \textbf{1.73} (0.48) & \textbf{0.96} (0.08)\\
        &  II  & CCA & 0.78 (0.10) & 1.52 (0.69) & 1.71 (0.59) & 0.67 (0.25)\\
        &    & NI & 0.84 (0.05) & 1.45 (0.38) & 1.47 (0.47) & 0.82 (0.06)\\
        &    & MI-Wood & 0.84 (0.05) & 1.44 (0.38) & 1.49 (0.48) & 0.81 (0.00)\\
        &    & MI-Bartlett & 0.84 (0.05) & 1.44 (0.38) & 1.49 (0.48) & 0.81 (0.02)\\
        &    & EG & \textbf{0.85} (0.05) & \textbf{0.95} (0.32) & \textbf{1.39} (0.46) & \textbf{0.99} (0.03)\\
        &  III  & CCA & 0.63 (0.12) & 2.02 (2.29) & 2.10 (0.54) & 0.41 (0.34)\\
        &    & NI & 0.78 (0.06) & 1.66 (0.50) & 1.76 (0.54) & \textbf{0.83} (0.06)\\
        &    & MI-Wood & 0.78 (0.06) & 1.64 (0.48) & 1.77 (0.52) & 0.81 (0.00)\\
        &    & MI-Bartlett & 0.78 (0.06) & 1.64 (0.47) & 1.77 (0.52) & 0.81 (0.02)\\
        &    & EG & \textbf{0.80} (0.06) & \textbf{1.00} (0.35) & \textbf{1.64} (0.58) & \textbf{0.83} (0.02)\\
    \midrule
    MAR & I   & CCA & 0.59 (0.11) & 1.39 (1.46) & 2.14 (0.59) & 0.20 (0.25)\\
        &    & NI & 0.76 (0.07) & 1.62 (0.57) & 1.71 (0.53) & 0.85 (0.09)\\
        &    & MI-Wood & 0.75 (0.07) & 1.66 (0.58) & 1.76 (0.52) & 0.81 (0.05)\\
        &    & MI-Bartlett & 0.75 (0.07) & 1.67 (0.62) & 1.76 (0.52) & 0.82 (0.06)\\
        &    & EG & \textbf{0.78} (0.06) & \textbf{0.92} (0.37) & \textbf{1.67} (0.54) & \textbf{0.96} (0.08)\\
         & II   & CCA & 0.81 (0.10) & 1.83 (1.99) & 1.43 (0.67) & 0.62 (0.25)\\
        &    & NI & 0.87 (0.05) & 1.48 (0.38) & 1.25 (0.51) & 0.82 (0.03)\\
        &    & MI-Wood & 0.87 (0.05) & 1.50 (0.41) & 1.26 (0.52) & 0.81 (0.00)\\
        &    & MI-Bartlett & 0.87 (0.05) & 1.49 (0.40) & 1.25 (0.51) & 0.81 (0.00)\\
        &    & EG & \textbf{0.88} (0.05) & \textbf{0.99} (0.38) & \textbf{1.18} (0.52) & \textbf{0.99} (0.03)\\
        & III   & CCA & 0.67 (0.14) & 1.62 (1.45) & 1.92 (0.64) & 0.45 (0.33)\\
        &    & NI & 0.80 (0.07) & 1.68 (0.50) & 1.68 (0.56) & 0.82 (0.05)\\
        &    & MI-Wood & 0.80 (0.07) & 1.66 (0.49) & 1.68 (0.55) & 0.81 (0.00)\\
        &    & MI-Bartlett & 0.80 (0.07) & 1.66 (0.49) & 1.68 (0.55) & 0.81 (0.00)\\
        &    & EG & \textbf{0.82} (0.06) & \textbf{0.96} (0.38) & \textbf{1.56} (0.62) & \textbf{0.83} (0.02)\\
    \midrule
     MARviol & I   & CCA & 0.58 (0.11) & 1.55 (1.39) & 2.17 (0.57) & 0.20 (0.24)\\
        &    & NI & 0.76 (0.06) & 1.63 (0.60) & 1.76 (0.54) & 0.84 (0.09)\\
        &    & MI-Wood & 0.75 (0.06) & 1.66 (0.60) & 1.79 (0.53) & 0.81 (0.05)\\
        &    & MI-Bartlett & 0.75 (0.07) & 1.65 (0.59) & 1.79 (0.54) & 0.83 (0.06)\\
        &    & EG & \textbf{0.78} (0.06) & \textbf{0.93} (0.37) & \textbf{1.69} (0.54) & \textbf{0.96} (0.08)\\
        &  II  & CCA & 0.78 (0.12) & 1.81 (1.10) & 1.55 (0.72) & 0.56 (0.28)\\
        &    & NI & 0.87 (0.05) & 1.45 (0.39) & 1.30 (0.55) & 0.81 (0.03)\\
        &    & MI-Wood & 0.87 (0.05) & 1.49 (0.41) & 1.30 (0.55) & 0.81 (0.00)\\
        &    & MI-Bartlett & 0.87 (0.05) & 1.48 (0.40) & 1.30 (0.55) & 0.81 (0.02)\\
        &    & EG & \textbf{0.88} (0.05) & \textbf{0.98} (0.42) & \textbf{1.22} (0.54) & \textbf{0.99} (0.03)\\
        &  III  & CCA & 0.65 (0.13) & 1.61 (2.09) & 1.98 (0.61) & 0.40 (0.33)\\
        &    & NI & 0.80 (0.07) & 1.69 (0.53) & 1.68 (0.55) & \textbf{0.83} (0.05)\\
        &    & MI-Wood & 0.79 (0.07) & 1.64 (0.50) & 1.67 (0.54) & 0.81 (0.00)\\
        &    & MI-Bartlett & 0.79 (0.07) & 1.64 (0.48) & 1.67 (0.54) & 0.81 (0.00)\\
        &    & EG & \textbf{0.82} (0.06) & \textbf{0.99} (0.45) & \textbf{1.55} (0.60) & \textbf{0.83} (0.02)\\
    \bottomrule
  \end{tabularx}
  \end{threeparttable}
  \end{center}
\end{table}

\pagebreak

\begin{table}[!ht]
  \begin{center}
  \caption{Simulation results for $r=0.15$ under various settings for missing mechanism. For each performance metric, the mean is reported with the standard deviation in parentheses. The best results are highlighted in boldface.}
\label{table:r0.15}
\begin{threeparttable}
  \fontsize{10}{11}\selectfont
  \begin{tabularx}{.87\textwidth}{lll| cccc}
    \toprule
    \multirow{2}{*}{Setting} & \multirow{2}{*}{Scenario}  &  \multirow{2}{*}{Method} &  \multirow{2}{*}{c-index}  &  Calibration  &  \multirow{2}{*}{IBS $\times 10^{1}$} & \multirow{2}{*}{MCC} \\
    & & & & slope  &  & \\
    \midrule
    MCAR &  I  & CCA & 0.63 (0.10) & 1.42 (1.29) & 2.05 (0.45) & 0.38 (0.29)\\
        &    & NI & 0.76 (0.05) & 1.66 (0.47) & 1.73 (0.42) & 0.89 (0.10)\\
        &    & MI-Wood & 0.74 (0.05) & 1.66 (0.49) & 1.81 (0.39) & 0.81 (0.05)\\
        &    & MI-Bartlett & 0.74 (0.05) & 1.65 (0.50) & 1.80 (0.39) & 0.85 (0.08)\\
        &    & EG & \textbf{0.77} (0.05) & \textbf{0.99} (0.29) & \textbf{1.66} (0.41) & \textbf{0.96} (0.08)\\
        & II   & CCA & 0.81 (0.07) & 1.48 (0.64) & 1.56 (0.48) & 0.79 (0.21)\\
        &    & NI & 0.84 (0.04) & 1.41 (0.28) & 1.44 (0.40) & 0.83 (0.06)\\
        &    & MI-Wood & 0.84 (0.04) & 1.42 (0.26) & 1.46 (0.40) & 0.81 (0.00)\\
        &    & MI-Bartlett & 0.84 (0.04) & 1.42 (0.26) & 1.46 (0.40) & 0.81 (0.03)\\
        &    & EG & \textbf{0.85} (0.04) & \textbf{0.98} (0.23) & \textbf{1.37} (0.41) & \textbf{0.99} (0.03)\\
        &  III  & CCA & 0.69 (0.09) & 1.65 (1.07) & 1.93 (0.43) & 0.62 (0.29)\\
        &    & NI & 0.78 (0.05) & 1.62 (0.41) & 1.70 (0.37) & \textbf{0.84} (0.08)\\
        &    & MI-Wood & 0.78 (0.05) & 1.61 (0.40) & 1.72 (0.35) & 0.81 (0.00)\\
        &    & MI-Bartlett & 0.78 (0.05) & 1.60 (0.40) & 1.72 (0.36) & 0.82 (0.03)\\
        &    & EG & \textbf{0.80} (0.05) & \textbf{0.99} (0.26) & \textbf{1.57} (0.41) & 0.83 (0.02)\\
    \midrule
    MAR  & I   & CCA & 0.64 (0.11) & 1.55 (1.19) & 2.01 (0.46) & 0.33 (0.27) \\
&& NI & 0.76 (0.06) & 1.65 (0.49) & 1.70 (0.41) & 0.85 (0.09) \\
&& MI-Wood & 0.75 (0.06) & 1.67 (0.52) & 1.73 (0.39) & 0.81 (0.05) \\
&& MI-Bartlett & 0.75 (0.06) & 1.67 (0.51) & 1.72 (0.39) & 0.84 (0.08) \\
&& EG & \textbf{0.79} (0.05) & \textbf{0.95} (0.32) & \textbf{1.61} (0.42) & \textbf{0.96} (0.08) \\
        &  II  & CCA & 0.84 (0.07) & 1.66 (0.95) & 1.31 (0.55) & 0.74 (0.22)\\
        &    & NI & 0.87 (0.04) & 1.44 (0.33) & 1.22 (0.46) & 0.82 (0.05)\\
        &    & MI-Wood & 0.87 (0.04) & 1.49 (0.37) & 1.23 (0.45) & 0.81 (0.00)\\
        &    & MI-Bartlett & 0.87 (0.04) & 1.46 (0.34) & 1.23 (0.45) & 0.82 (0.03)\\
        &    & EG & \textbf{0.88} (0.04) & \textbf{0.97} (0.28) & \textbf{1.14} (0.44) & \textbf{0.99} (0.03)\\
        & III   & CCA & 0.71 (0.11) & 1.80 (2.28) & 1.85 (0.56) & 0.60 (0.32)\\
        &    & NI & 0.80 (0.05) & 1.62 (0.39) & 1.62 (0.50) & 0.82 (0.06)\\
        &    & MI-Wood & 0.80 (0.05) & 1.67 (0.41) & 1.62 (0.49) & 0.81 (0.00)\\
        &    & MI-Bartlett & 0.80 (0.05) & 1.63 (0.38) & 1.62 (0.49) & 0.82 (0.04)\\
        &    & EG & \textbf{0.82} (0.05) & \textbf{1.00} (0.32) & \textbf{1.50} (0.52) & \textbf{0.83} (0.02)\\
    \midrule
     MARviol &  I  & CCA & 0.63 (0.11) & 1.43 (0.92) & 1.99 (0.42) & 0.31 (0.27)\\
        &    & NI & 0.76 (0.06) & 1.67 (0.51) & 1.70 (0.42) & 0.85 (0.10)\\
        &    & MI-Wood & 0.75 (0.06) & 1.68 (0.52) & 1.74 (0.40) & 0.82 (0.05)\\
        &    & MI-Bartlett & 0.75 (0.06) & 1.71 (0.55) & 1.73 (0.41) & 0.84 (0.09)\\
        &    & EG & \textbf{0.79} (0.05) & \textbf{0.96} (0.32) & \textbf{1.62} (0.43) & \textbf{0.96} (0.08)\\
        &  II  & CCA & 0.83 (0.08) & 1.59 (0.71) & 1.36 (0.58) & 0.71 (0.24)\\
        &    & NI & 0.87 (0.04) & 1.45 (0.32) & 1.23 (0.45) & 0.83 (0.06)\\
        &    & MI-Wood & 0.87 (0.04) & 1.47 (0.33) & 1.23 (0.44) & 0.81 (0.00)\\
        &    & MI-Bartlett & 0.87 (0.04) & 1.47 (0.33) & 1.24 (0.44) & 0.81 (0.02)\\
        &    & EG & \textbf{0.88} (0.04) & \textbf{0.97} (0.28) & \textbf{1.15} (0.42) & \textbf{0.99} (0.03)\\
        & III   & CCA & 0.71 (0.12) & 1.87 (1.79) & 1.86 (0.50) & 0.59 (0.32)\\
        &    & NI & 0.80 (0.05) & 1.63 (0.39) & 1.62 (0.49) & 0.82 (0.06)\\
        &    & MI-Wood & 0.80 (0.05) & 1.66 (0.40) & 1.62 (0.48) & 0.81 (0.00)\\
        &    & MI-Bartlett & 0.80 (0.05) & 1.66 (0.40) & 1.62 (0.48) & 0.81 (0.02)\\
        &    & EG & \textbf{0.82} (0.04) & \textbf{1.01} (0.32) & \textbf{1.51} (0.53) & \textbf{0.83} (0.02)\\
    \bottomrule
  \end{tabularx}
  \end{threeparttable}
  \end{center}
\end{table}

\pagebreak

\begin{table}[!ht]
  \begin{center}
  \caption{Simulation results for $r=0.25$ under various settings for missing mechanism. For each performance metric, the mean is reported with the standard deviation in parentheses. The best results are highlighted in boldface.}
\label{table:r0.25}
\begin{threeparttable}
  \fontsize{10}{11}\selectfont
  \begin{tabularx}{.87\textwidth}{lll| cccc}
    \toprule
    \multirow{2}{*}{Setting \hspace{1.5em}} & \multirow{2}{*}{Scenario}  &  \multirow{2}{*}{Method} &  \multirow{2}{*}{c-index}  &  Calibration  &  \multirow{2}{*}{IBS $\times 10^{1}$} & \multirow{2}{*}{MCC} \\
    & & & & slope  &  & \\
    \midrule
    MCAR & I  & CCA & 0.70 (0.08) & 1.61 (1.65) & 1.87 (0.37) & 0.60 (0.27)\\
        &    & NI & 0.77 (0.04) & 1.65 (0.42) & 1.68 (0.35) & 0.93 (0.10)\\
        &    & MI-Wood & 0.74 (0.04) & 1.64 (0.39) & 1.78 (0.34) & 0.82 (0.07)\\
        &    & MI-Bartlett & 0.75 (0.04) & 1.62 (0.42) & 1.77 (0.34) & 0.88 (0.10)\\
        &    & EG & \textbf{0.78} (0.04) & \textbf{1.00} (0.24) & \textbf{1.63} (0.36) & \textbf{0.96} (0.08)\\
        &  II  & CCA & 0.84 (0.03) & 1.48 (0.32) & 1.42 (0.46) & 0.91 (0.13)\\
        &    & NI & 0.84 (0.03) & 1.36 (0.20) & 1.42 (0.45) & 0.86 (0.08)\\
        &    & MI-Wood & 0.84 (0.03) & 1.41 (0.21) & 1.44 (0.44) & 0.81 (0.02)\\
        &    & MI-Bartlett & 0.84 (0.03) & 1.39 (0.19) & 1.44 (0.44) & 0.83 (0.06)\\
        &    & EG & \textbf{0.85} (0.03) & \textbf{0.98} (0.16) & \textbf{1.36} (0.39) & \textbf{0.99} (0.03)\\
        &  III  & CCA & 0.76 (0.05) & 1.63 (1.13) & 1.68 (0.41) & \textbf{0.87} (0.16)\\
        &    & NI & 0.78 (0.03) & 1.55 (0.32) & 1.62 (0.35) & \textbf{0.87} (0.09)\\
        &    & MI-Wood & 0.78 (0.03) & 1.60 (0.29) & 1.66 (0.33) & 0.81 (0.02)\\
        &    & MI-Bartlett & 0.78 (0.03) & 1.58 (0.30) & 1.66 (0.33) & 0.83 (0.06)\\
        &    & EG & \textbf{0.80} (0.03) & \textbf{1.00} (0.20) & \textbf{1.47} (0.35) & 0.83 (0.02)\\
    \midrule
    MAR &  I & CCA & 0.70 (0.08) & 1.58 (1.21) & 1.88 (0.38) & 0.55 (0.27)\\
        &    & NI & 0.76 (0.05) & 1.62 (0.38) & 1.68 (0.35) & 0.89 (0.10)\\
        &    & MI-Wood & 0.75 (0.05) & 1.66 (0.40) & 1.74 (0.34) & 0.82 (0.07)\\
        &    & MI-Bartlett & 0.75 (0.04) & 1.64 (0.39) & 1.73 (0.35) & 0.86 (0.09)\\
        &    & EG & \textbf{0.78} (0.04) & \textbf{0.98} (0.22) & \textbf{1.57} (0.35) & \textbf{0.96} (0.08)\\
        &  II  & CCA & 0.85 (0.05) & 1.48 (0.39) & 1.26 (0.46) & 0.86 (0.16)\\
        &    & NI & 0.86 (0.03) & 1.42 (0.20) & 1.23 (0.40) & 0.83 (0.06)\\
        &    & MI-Wood & 0.86 (0.03) & 1.44 (0.21) & 1.24 (0.40) & 0.81 (0.00)\\
        &    & MI-Bartlett & 0.86 (0.03) & 1.43 (0.21) & 1.24 (0.40) & 0.82 (0.04)\\
        &    & EG & \textbf{0.87} (0.03) & \textbf{0.99} (0.17) & \textbf{1.16} (0.35) & \textbf{0.99} (0.03)\\
        &  III  & CCA & 0.77 (0.06) & 1.57 (0.77) & 1.65 (0.48) & 0.83 (0.20)\\
        &    & NI & 0.79 (0.04) & 1.60 (0.35) & 1.57 (0.43) & \textbf{0.84} (0.07)\\
        &    & MI-Wood & 0.79 (0.04) & 1.63 (0.32) & 1.59 (0.42) & 0.81 (0.00)\\
        &    & MI-Bartlett & 0.79 (0.04) & 1.61 (0.33) & 1.59 (0.42) & 0.82 (0.04)\\
        &    & EG & \textbf{0.82} (0.04) & \textbf{1.00} (0.22) & \textbf{1.43} (0.43) & 0.83 (0.02)\\
    \midrule
     MARviol & I   & CCA & 0.70 (0.08) & 1.50 (0.81) & 1.90 (0.40) & 0.57 (0.24)\\
        &    & NI & 0.76 (0.05) & 1.61 (0.42) & 1.67 (0.36) & 0.89 (0.10)\\
        &    & MI-Wood & 0.75 (0.04) & 1.65 (0.42) & 1.74 (0.34) & 0.81 (0.06)\\
        &    & MI-Bartlett & 0.75 (0.04) & 1.65 (0.45) & 1.72 (0.35) & 0.88 (0.09)\\
        &    & EG & \textbf{0.78} (0.04) & \textbf{0.98} (0.22) & \textbf{1.58} (0.36) & \textbf{0.96} (0.08)\\
        & II   & CCA & 0.85 (0.04) & 1.52 (0.59) & 1.27 (0.43) & 0.87 (0.14)\\
        &    & NI & 0.86 (0.03) & 1.39 (0.19) & 1.24 (0.41) & 0.83 (0.06)\\
        &    & MI-Wood & 0.86 (0.03) & 1.42 (0.21) & 1.26 (0.40) & 0.81 (0.00)\\
        &    & MI-Bartlett & 0.86 (0.03) & 1.41 (0.21) & 1.26 (0.40) & 0.83 (0.05)\\
        &    & EG & \textbf{0.87} (0.03) & \textbf{0.98} (0.17) & \textbf{1.18} (0.35) & \textbf{0.99} (0.03)\\
        & III    & CCA & 0.77 (0.06) & 1.66 (1.01) & 1.66 (0.48) & 0.82 (0.19)\\
        &    & NI & 0.79 (0.04) & 1.58 (0.34) & 1.57 (0.43) & \textbf{0.83} (0.07)\\
        &    & MI-Wood & 0.79 (0.04) & 1.62 (0.31) & 1.59 (0.41) & 0.81 (0.00)\\
        &    & MI-Bartlett & 0.79 (0.04) & 1.59 (0.33) & 1.59 (0.41) & 0.82 (0.05)\\
        &    & EG & \textbf{0.82} (0.04) & \textbf{1.00} (0.22) & \textbf{1.44} (0.44) & \textbf{0.83} (0.02)\\
    \bottomrule
  \end{tabularx}
  \end{threeparttable}
  \end{center}
\end{table}

\section{}

Additional simulations are conducted to evaluate our proposed method in comparison to other methods with a continuous missing covariate. The data-generating distribution and setup resembles Section 3.1 except for $\bb V$, where $\bb V \sim N(0.4|U_1| - 0.1, 0.2^2)$, and the types of models to be used for imputing variables, including mean imputation for NI and the default imputation methods for MI approaches. Under this simulation design, the results remained largely the same as in the binary missing covariate case, demonstrating that our proposed method outperformed its alternatives in terms of a higher c-index, calibration slope closer to 1, lower IBS, and higher MCC in a majority of cases (Web Tables \ref{table:MCAR:cont}--\ref{table:MARviol:cont}).
Specifically, our proposed EG method outperformed its competitors in terms of achieving a higher c-index, better calibration, and higher MCC, along with a lower standard deviation. 
In contrast, the CCA method had a low discriminatory ability, and the other methods, such as NI and the two MI approaches, showed poor calibration, indicating inaccurate risk estimates. 
The MI-Bartlett method performed slightly better than the MI-Wood method in terms of c-index, IBS, and MCC. 
However, the MI-Bartlett method experienced overfitting, characterized by a calibration slope far from 1 with increased variability.
The superiority of the proposed EG method was most evident in Scenario I, followed by a lesser, yet still notable, benefit in Scenario II. In Scenario III, the MI-Bartlett method showed a similar c-index to the EG method, with better IBS and/or MCC; this is expected, as the first two variables of $\bb U$ were included in the model based on domain knowledge, making it less suitable for the proposed method when $U_2$ had no effect. Nevertheless, the proposed EG method consistently yielded the best calibration, with significantly lower variability, across all scenarios by incorporating the prognostic index and making effective utilization of two-phase data.

\begin{table}[!ht]
  \begin{center}
\caption{Simulation results under the MCAR setting for a continuous missing covariate. For each performance metric, the mean is reported with the standard deviation in parentheses. The best results are highlighted in boldface.}
\label{table:MCAR:cont}
\begin{threeparttable}
  \fontsize{10}{11}\selectfont
  \begin{tabularx}{.76\textwidth}{ll| cccc}
    \toprule
    \multirow{2}{*}{Scenario} & \multirow{2}{*}{Method} &  \multirow{2}{*}{c-index}  &  Calibration  &  \multirow{2}{*}{IBS $\times 10^{1}$} & \multirow{2}{*}{MCC} \\
    & & & slope  &  & \\
    \midrule
    I   & CCA & 0.56 (0.11) & 1.56 (3.63) & 2.38 (0.70) & 0.15 (0.22) \\
& NI & 0.61 (0.12) & 2.36 (8.69) & 2.36 (0.90) & 0.33 (0.25) \\
& MI-Wood & 0.60 (0.12) & 1.83 (2.23) & 2.36 (0.84) & 0.32 (0.27) \\
& MI-Bartlett & 0.62 (0.12) & 3.30 (5.52) & 2.33 (0.90) & 0.39 (0.26) \\
& EG & \textbf{0.68} (0.11) & \textbf{0.78} (1.12) & \textbf{2.16} (0.86) & \textbf{0.65} (0.13) \\
    II  & CCA & 0.71 (0.14) & 1.67 (1.44) & 2.12 (0.75) & 0.44 (0.30) \\
& NI & 0.83 (0.07) & 1.59 (0.81) & 1.68 (0.67) & 0.79 (0.16) \\
& MI-Wood & 0.83 (0.08) & 1.80 (0.91) & 1.66 (0.70) & 0.71 (0.14) \\
& MI-Bartlett & 0.83 (0.07) & 1.82 (1.03) & 1.65 (0.70) & 0.78 (0.15) \\
& EG & \textbf{0.84} (0.07) & \textbf{0.93} (0.39) & \textbf{1.64} (0.70) & \textbf{0.87} (0.09) \\
    III & CCA & 0.61 (0.13) & 1.40 (1.46) & 2.31 (0.74) & 0.25 (0.29) \\
& NI & 0.72 (0.13) & 3.01 (10.6) & 2.09 (0.95) & 0.65 (0.28) \\
& MI-Wood & 0.72 (0.12) & 2.72 (6.26) & \textbf{2.02} (0.91) & 0.63 (0.24) \\
& MI-Bartlett & \textbf{0.73} (0.12) & 3.50 (8.95) & \textbf{2.02} (0.93) & \textbf{0.68} (0.24) \\
& EG & \textbf{0.73} (0.10) & \textbf{1.07} (2.34) & 2.06 (0.85) & 0.61 (0.19) \\
    \bottomrule
  \end{tabularx}
  \end{threeparttable}
  \end{center}
\end{table}

\begin{table}[!ht]
  \begin{center}
  \caption{Simulation results under the MAR setting for a continuous missing covariate. For each performance metric, the mean is reported with the standard deviation in parentheses. The best results are highlighted in boldface.}
\label{table:MAR:cont}
\begin{threeparttable}
  \fontsize{10}{11}\selectfont
  \begin{tabularx}{.76\textwidth}{ll| cccc}
    \toprule
    \multirow{2}{*}{Scenario} & \multirow{2}{*}{Method} &  \multirow{2}{*}{c-index}  &  Calibration  &  \multirow{2}{*}{IBS $\times 10^{1}$} & \multirow{2}{*}{MCC} \\
    & & & slope  &  & \\
    \midrule
    I   & CCA & 0.55 (0.09) & 1.25 (2.18) & 2.34 (0.67) & 0.11 (0.17) \\
& NI & 0.62 (0.13) & 1.63 (1.95) & 2.15 (0.77) & 0.34 (0.28) \\
& MI-Wood & 0.61 (0.12) & 2.08 (2.24) & 2.21 (0.75) & 0.32 (0.26) \\
& MI-Bartlett & 0.62 (0.13) & 1.18 (15.9) & 2.24 (0.76) & 0.38 (0.26) \\
& EG & \textbf{0.68} (0.12) & \textbf{0.85} (0.73) & \textbf{2.07} (0.73) & \textbf{0.65} (0.13) \\
    II  & CCA & 0.75 (0.16) & 1.88 (1.43) & 1.85 (0.94) & 0.45 (0.27) \\
& NI & 0.84 (0.09) & 1.67 (1.02) & 1.56 (0.95) & 0.74 (0.14) \\
& MI-Wood & 0.84 (0.09) & 1.85 (1.26) & 1.57 (0.97) & 0.72 (0.12) \\
& MI-Bartlett & 0.84 (0.09) & 1.82 (1.19) & 1.56 (0.97) & 0.78 (0.14) \\
& EG & \textbf{0.85} (0.08) & \textbf{0.96} (0.52) & \textbf{1.54} (0.88) & \textbf{0.89} (0.09) \\
    III & CCA & 0.62 (0.14) & 1.59 (1.32) & 2.12 (0.73) & 0.27 (0.31)\\
& NI & 0.71 (0.12) & 1.94 (1.66) & \textbf{1.99} (0.93) & 0.64 (0.25)\\
& MI-Wood & 0.72 (0.12) & 2.37 (3.28) & 2.03 (1.11) & 0.65 (0.20)\\
& MI-Bartlett & 0.72 (0.12) & 2.18 (2.26) & \textbf{1.99} (1.10) & \textbf{0.69} (0.21)\\
& EG & \textbf{0.73} (0.12) & \textbf{0.91} (1.10) & 2.07 (0.98) & 0.61 (0.18)\\
    \bottomrule
  \end{tabularx}
  \end{threeparttable}
  \end{center}
\end{table}

\begin{table}[!ht]
  \begin{center}
  \caption{Simulation results under the MAR setting with a mild-to-moderate violation for a continuous missing covariate. For each performance metric, the mean is reported with the standard deviation in parentheses. The best results are highlighted in boldface.}
\label{table:MARviol:cont}
\begin{threeparttable}
  \fontsize{10}{11}\selectfont
  \begin{tabularx}{.76\textwidth}{ll| cccc}
    \toprule
    \multirow{2}{*}{Scenario} & \multirow{2}{*}{Method} &  \multirow{2}{*}{c-index}  &  Calibration  &  \multirow{2}{*}{IBS $\times 10^{1}$} & \multirow{2}{*}{MCC} \\
    & & & slope  &  & \\
    \midrule
    I   & CCA & 0.55 (0.10) & \textbf{1.09} (1.96) & 2.35 (0.66) & 0.13 (0.18) \\
& NI & 0.62 (0.13) & 1.62 (2.48) & 2.16 (0.76) & 0.33 (0.27) \\
& MI-Wood & 0.60 (0.12) & 2.38 (4.22) & 2.23 (0.73) & 0.32 (0.25) \\
& MI-Bartlett & 0.62 (0.12) & 3.52 (6.93) & 2.24 (0.77) & 0.39 (0.28) \\
& EG & \textbf{0.68} (0.12) & 0.84 (0.70) & \textbf{2.06} (0.71) & \textbf{0.65} (0.13) \\
    II  & CCA & 0.76 (0.15) & 1.93 (1.65) & 1.85 (0.95) & 0.47 (0.26) \\
& NI & 0.84 (0.08) & 1.67 (1.02) & 1.55 (0.94) & 0.75 (0.14) \\
& MI-Wood & 0.84 (0.09) & 1.78 (1.19) & 1.55 (0.96) & 0.73 (0.12) \\
& MI-Bartlett & \textbf{0.85} (0.09) & 1.80 (1.15) & 1.55 (0.96) & 0.77 (0.14) \\
& EG & \textbf{0.85} (0.08) & \textbf{0.95} (0.50) & \textbf{1.52} (0.87) & \textbf{0.90} (0.09) \\
    III & CCA & 0.61 (0.14) & 1.57 (1.85) & 2.15 (0.73) & 0.27 (0.31) \\
& NI & 0.71 (0.12) & 1.89 (1.59) & \textbf{2.00} (0.92) & 0.64 (0.25) \\
& MI-Wood & 0.72 (0.11) & 2.18 (2.33) & 2.05 (1.10) & 0.65 (0.20) \\
& MI-Bartlett & \textbf{0.73} (0.12) & 2.22 (2.44) & 2.02 (1.10) & \textbf{0.69} (0.22) \\
& \textbf{EG} & \textbf{0.73} (0.12) & \textbf{0.89} (0.95) & 2.08 (0.97) & 0.61 (0.18) \\
    \bottomrule
  \end{tabularx}
  \end{threeparttable}
  \end{center}
\end{table}

\section{}

Additional simulations are conducted to evaluate our proposed method in comparison to other methods with two missing covariates. The data-generating distribution and setup resembles Sections 3.1, except for $\bb{V} = (V_1, V_2)$, where $V_1$ follows Section 3.1, $V_2$ follows Web Appendix D. Additionally, the probability of entire $\bb V$ being missing is determined based on $V_1$ for MAR with a mild-to-moderate violation.
Web Tables \ref{table:MCAR:two}--\ref{table:MARviol:two} report the performance metrics for different missing mechanisms under this simulation design. The results remained largely consistent, with our proposed method outperforming its alternatives in terms of a higher c-index, a calibration slope closer to 1, lower IBS, and higher MCC in the majority of cases.

\begin{table}[!ht]
  \begin{center}
\caption{Simulation results under the MCAR setting for two missing covariates. For each performance metric, the mean is reported with the standard deviation in parentheses. The best results are highlighted in boldface.}
\label{table:MCAR:two}
\begin{threeparttable}
  \fontsize{10}{11}\selectfont
  \begin{tabularx}{.76\textwidth}{ll| cccc}
    \toprule
    \multirow{2}{*}{Scenario} & \multirow{2}{*}{Method} &  \multirow{2}{*}{c-index}  &  Calibration  &  \multirow{2}{*}{IBS $\times 10^{1}$} & \multirow{2}{*}{MCC} \\
    & & & slope  &  & \\
    \midrule
    I   & CCA & 0.56 (0.11) & 1.56 (3.63) & 2.38 (0.70) & 0.15 (0.22) \\
& NI & 0.61 (0.12) & 2.36 (8.69) & 2.36 (0.90) & 0.33 (0.25) \\
& MI-Wood & 0.60 (0.12) & 1.83 (2.23) & 2.36 (0.84) & 0.32 (0.27) \\
& MI-Bartlett & 0.62 (0.12) & 3.30 (5.52) & 2.33 (0.90) & 0.39 (0.26) \\
& EG & \textbf{0.68} (0.11) & \textbf{0.78} (1.12) & \textbf{2.16} (0.86) & \textbf{0.65} (0.13) \\
    II  & CCA & 0.71 (0.14) & 1.67 (1.44) & 2.12 (0.75) & 0.44 (0.30) \\
& NI & 0.83 (0.07) & 1.59 (0.81) & 1.68 (0.67) & 0.79 (0.16) \\
& MI-Wood & 0.83 (0.08) & 1.80 (0.91) & 1.66 (0.70) & 0.71 (0.14) \\
& MI-Bartlett & 0.83 (0.07) & 1.82 (1.03) & 1.65 (0.70) & 0.78 (0.15) \\
& EG & \textbf{0.84} (0.07) & \textbf{0.93} (0.39) & \textbf{1.64} (0.70) & \textbf{0.87} (0.09) \\
    III & CCA & 0.61 (0.13) & 1.40 (1.46) & 2.31 (0.74) & 0.25 (0.29) \\
& NI & 0.72 (0.13) & 3.01 (10.6) & 2.09 (0.95) & 0.65 (0.28) \\
& MI-Wood & 0.72 (0.12) & 2.72 (6.26) & \textbf{2.02} (0.91) & 0.63 (0.24) \\
& MI-Bartlett & \textbf{0.73} (0.12) & 3.50 (8.95) & \textbf{2.02} (0.93) & \textbf{0.68} (0.24) \\
& EG & \textbf{0.73} (0.10) & \textbf{1.07} (2.34) & 2.06 (0.85) & 0.61 (0.19) \\
    \bottomrule
  \end{tabularx}
  \end{threeparttable}
  \end{center}
\end{table}

\begin{table}[!ht]
  \begin{center}
  \caption{Simulation results under the MAR setting for two missing covariates. For each performance metric, the mean is reported with the standard deviation in parentheses. The best results are highlighted in boldface.}
\label{table:MAR:two}
\begin{threeparttable}
  \fontsize{10}{11}\selectfont
  \begin{tabularx}{.76\textwidth}{ll| cccc}
    \toprule
    \multirow{2}{*}{Scenario} & \multirow{2}{*}{Method} &  \multirow{2}{*}{c-index}  &  Calibration  &  \multirow{2}{*}{IBS $\times 10^{1}$} & \multirow{2}{*}{MCC} \\
    & & & slope  &  & \\
    \midrule
    I   & CCA & 0.55 (0.09) & 1.25 (2.18) & 2.34 (0.67) & 0.11 (0.17) \\
& NI & 0.62 (0.13) & 1.63 (1.95) & 2.15 (0.77) & 0.34 (0.28) \\
& MI-Wood & 0.61 (0.12) & 2.08 (2.24) & 2.21 (0.75) & 0.32 (0.26) \\
& MI-Bartlett & 0.62 (0.13) & 1.18 (15.9) & 2.24 (0.76) & 0.38 (0.26) \\
& EG & \textbf{0.68} (0.12) & \textbf{0.85} (0.73) & \textbf{2.07} (0.73) & \textbf{0.65} (0.13) \\
    II  & CCA & 0.75 (0.16) & 1.88 (1.43) & 1.85 (0.94) & 0.45 (0.27) \\
& NI & 0.84 (0.09) & 1.67 (1.02) & 1.56 (0.95) & 0.74 (0.14) \\
& MI-Wood & 0.84 (0.09) & 1.85 (1.26) & 1.57 (0.97) & 0.72 (0.12) \\
& MI-Bartlett & 0.84 (0.09) & 1.82 (1.19) & 1.56 (0.97) & 0.78 (0.14) \\
& EG & \textbf{0.85} (0.08) & \textbf{0.96} (0.52) & \textbf{1.54} (0.88) & \textbf{0.89} (0.09) \\
    III & CCA & 0.62 (0.14) & 1.59 (1.32) & 2.12 (0.73) & 0.27 (0.31)\\
& NI & 0.71 (0.12) & 1.94 (1.66) & \textbf{1.99} (0.93) & 0.64 (0.25)\\
& MI-Wood & 0.72 (0.12) & 2.37 (3.28) & 2.03 (1.11) & 0.65 (0.20)\\
& MI-Bartlett & 0.72 (0.12) & 2.18 (2.26) & \textbf{1.99} (1.10) & \textbf{0.69} (0.21)\\
& EG & \textbf{0.73} (0.12) & \textbf{0.91} (1.10) & 2.07 (0.98) & 0.61 (0.18)\\
    \bottomrule
  \end{tabularx}
  \end{threeparttable}
  \end{center}
\end{table}

\begin{table}[!ht]
  \begin{center}
  \caption{Simulation results under the MAR setting with a mild-to-moderate violation for two missing covariates. For each performance metric, the mean is reported with the standard deviation in parentheses. The best results are highlighted in boldface.}
\label{table:MARviol:two}
\begin{threeparttable}
  \fontsize{10}{11}\selectfont
  \begin{tabularx}{.76\textwidth}{ll| cccc}
    \toprule
    \multirow{2}{*}{Scenario} & \multirow{2}{*}{Method} &  \multirow{2}{*}{c-index}  &  Calibration  &  \multirow{2}{*}{IBS $\times 10^{1}$} & \multirow{2}{*}{MCC} \\
    & & & slope  &  & \\
    \midrule
    I   & CCA & 0.55 (0.10) & \textbf{1.09} (1.96) & 2.35 (0.66) & 0.13 (0.18) \\
& NI & 0.62 (0.13) & 1.62 (2.48) & 2.16 (0.76) & 0.33 (0.27) \\
& MI-Wood & 0.60 (0.12) & 2.38 (4.22) & 2.23 (0.73) & 0.32 (0.25) \\
& MI-Bartlett & 0.62 (0.12) & 3.52 (6.93) & 2.24 (0.77) & 0.39 (0.28) \\
& EG & \textbf{0.68} (0.12) & 0.84 (0.70) & \textbf{2.06} (0.71) & \textbf{0.65} (0.13) \\
    II  & CCA & 0.76 (0.15) & 1.93 (1.65) & 1.85 (0.95) & 0.47 (0.26) \\
& NI & 0.84 (0.08) & 1.67 (1.02) & 1.55 (0.94) & 0.75 (0.14) \\
& MI-Wood & 0.84 (0.09) & 1.78 (1.19) & 1.55 (0.96) & 0.73 (0.12) \\
& MI-Bartlett & \textbf{0.85} (0.09) & 1.80 (1.15) & 1.55 (0.96) & 0.77 (0.14) \\
& EG & \textbf{0.85} (0.08) & \textbf{0.95} (0.50) & \textbf{1.52} (0.87) & \textbf{0.90} (0.09) \\
    III & CCA & 0.61 (0.14) & 1.57 (1.85) & 2.15 (0.73) & 0.27 (0.31) \\
& NI & 0.71 (0.12) & 1.89 (1.59) & \textbf{2.00} (0.92) & 0.64 (0.25) \\
& MI-Wood & 0.72 (0.11) & 2.18 (2.33) & 2.05 (1.10) & 0.65 (0.20) \\
& MI-Bartlett & \textbf{0.73} (0.12) & 2.22 (2.44) & 2.02 (1.10) & \textbf{0.69} (0.22) \\
& \textbf{EG} & \textbf{0.73} (0.12) & \textbf{0.89} (0.95) & 2.08 (0.97) & 0.61 (0.18) \\
    \bottomrule
  \end{tabularx}
  \end{threeparttable}
  \end{center}
\end{table}

\section{}

Additional simulations are conducted to investigate the impact of the violation of the proportional hazards assumption on the performance of our method compared to other existing methods. The data-generating distribution and setup resembles Sections 3.1, except that we introduce time-dependent effect of $\bb U$ on the survival times. 
Web Tables \ref{table:MCAR:PH}--\ref{table:MARviol:PH} show the performance metrics for different missing mechanisms under the non-proportionality of the hazards. The results showed a reduced c-index and biased risk estimates across all methods, as expected, although the overall impact on our method was relatively small. For example, the c-index reduction in Scenarios II and III was considerably smaller for our method compared to the other methods. In addition, the variable selection performance, demonstrated by MCC, was most significantly impacted for all methods except our method. Our proposed method experienced only a minimal decrease in MCC, due to the benefit of incorporating domain knowledge and prognostic index. \\

\begin{table}[!ht]
  \begin{center}
\caption{Simulation results under the MCAR setting when the proportional hazards assumption is violated. For each performance metric, the mean is reported with the standard deviation in parentheses. The best results are highlighted in boldface.}
\label{table:MCAR:PH}
\begin{threeparttable}
  \fontsize{10}{11}\selectfont
  \begin{tabularx}{.76\textwidth}{ll| cccc}
    \toprule
    \multirow{2}{*}{Scenario} & \multirow{2}{*}{Method} &  \multirow{2}{*}{c-index}  &  Calibration  &  \multirow{2}{*}{IBS $\times 10^{1}$} & \multirow{2}{*}{MCC} \\
    & & & slope  &  & \\
    \midrule
    I   & CCA & 0.53 (0.09) & 1.18 (2.58) & 2.24 (0.68) & 0.08 (0.18) \\
& NI & 0.52 (0.07) & 0.25 (9.87) & 2.19 (0.64) & 0.05 (0.16) \\
& MI-Wood & 0.51 (0.04) & 1.49 (1.87) & 2.17 (0.60) & 0.03 (0.11) \\
& MI-Bartlett & 0.55 (0.10) & 8.81 (19.1) & 2.19 (0.63) & 0.12 (0.19) \\
& EG & \textbf{0.66} (0.12) & \textbf{0.86} (1.04) & \textbf{2.05} (0.72) & \textbf{0.57} (0.06) \\
    II  & CCA & 0.56 (0.12) & \textbf{0.86} (5.35) & 2.11 (0.69) & 0.19 (0.28) \\
& NI & 0.60 (0.14) & 2.09 (2.36) & 2.01 (0.63) & 0.31 (0.31) \\
& MI-Wood & 0.58 (0.12) & 2.14 (3.27) & 2.06 (0.65) & 0.27 (0.29) \\
& MI-Bartlett & 0.61 (0.13) & 2.90 (6.08) & 2.07 (0.69) & 0.38 (0.28) \\
& EG & \textbf{0.74} (0.12) & 1.24 (2.58) & \textbf{1.80} (0.76) & \textbf{0.82} (0.05) \\
    III & CCA & 0.53 (0.10) & \textbf{1.12} (2.79) & 2.22 (0.67) & 0.11 (0.23) \\
& NI & 0.53 (0.09) & 4.35 (11.4) & 2.14 (0.57) & 0.13 (0.25) \\
& MI-Wood & 0.52 (0.08) & 1.59 (3.42) & 2.17 (0.60) & 0.10 (0.21) \\
& MI-Bartlett & 0.56 (0.11) & 7.67 (28.2) & 2.16 (0.66) & 0.24 (0.25) \\
& EG & \textbf{0.65} (0.14) & 0.82 (0.95) & \textbf{2.00} (0.67) & \textbf{0.42} (0.11) \\
    \bottomrule
  \end{tabularx}
  \end{threeparttable}
  \end{center}
\end{table}

\begin{table}[!ht]
  \begin{center}
  \caption{Simulation results under the MAR setting when the proportional hazards assumption is violated. For each performance metric, the mean is reported with the standard deviation in parentheses. The best results are highlighted in boldface.}
\label{table:MAR:PH}
\begin{threeparttable}
  \fontsize{10}{11}\selectfont
  \begin{tabularx}{.76\textwidth}{ll| cccc}
    \toprule
    \multirow{2}{*}{Scenario} & \multirow{2}{*}{Method} &  \multirow{2}{*}{c-index}  &  Calibration  &  \multirow{2}{*}{IBS $\times 10^{1}$} & \multirow{2}{*}{MCC} \\
    & & & slope  &  & \\
    \midrule
    I   & CCA & 0.53 (0.09) & \textbf{0.98} (9.14) & 2.18 (0.67) & 0.07 (0.18) \\
& NI & 0.51 (0.06) & 1.73 (12.3) & 2.08 (0.53) & 0.05 (0.14) \\
& MI-Wood & 0.50 (0.04) & 1.19 (2.99) & 2.07 (0.51) & 0.01 (0.07) \\
& MI-Bartlett & 0.54 (0.09) & 2.61 (8.14) & 2.10 (0.64) & 0.11 (0.18) \\
& EG & \textbf{0.66} (0.13) & 0.86 (0.90) & \textbf{2.03} (0.84) & \textbf{0.57} (0.06) \\
    II  & CCA & 0.58 (0.13) & \textbf{1.44} (8.63) & 1.88 (0.88) & 0.21 (0.30) \\
& NI & 0.61 (0.15) & 2.60 (2.81) & 1.78 (0.68) & 0.27 (0.31) \\
& MI-Wood & 0.60 (0.14) & 2.12 (2.55) & 1.79 (0.69) & 0.25 (0.29) \\
& MI-Bartlett & 0.65 (0.15) & 5.82 (8.49) & 1.71 (0.71) & 0.37 (0.31) \\
& EG & \textbf{0.77} (0.13) & 2.37 (10.9) & \textbf{1.56} (0.76) & \textbf{0.82} (0.05) \\
    III & CCA & 0.55 (0.11) & 1.52 (3.65) & 2.00 (0.63) & 0.14 (0.26) \\
& NI & 0.54 (0.11) & 3.77 (7.98) & 1.98 (0.60) & 0.15 (0.27) \\
& MI-Wood & 0.53 (0.09) & 2.66 (4.33) & 1.94 (0.53) & 0.08 (0.19) \\
& MI-Bartlett & 0.57 (0.11) & 11.4 (40.8) & 1.96 (0.61) & 0.25 (0.28) \\
& EG & \textbf{0.67} (0.16) & \textbf{1.04} (2.16) & \textbf{1.86} (0.68) & \textbf{0.42} (0.11) \\
    \bottomrule
  \end{tabularx}
  \end{threeparttable}
  \end{center}
\end{table}

\begin{table}[!ht]
  \begin{center}
  \caption{Simulation results under the MAR setting with a mild-to-moderate violation when the proportional hazards assumption is violated. For each performance metric, the mean is reported with the standard deviation in parentheses. The best results are highlighted in boldface.}
\label{table:MARviol:PH}
\begin{threeparttable}
  \fontsize{10}{11}\selectfont
  \begin{tabularx}{.76\textwidth}{ll| cccc}
    \toprule
    \multirow{2}{*}{Scenario} & \multirow{2}{*}{Method} &  \multirow{2}{*}{c-index}  &  Calibration  &  \multirow{2}{*}{IBS $\times 10^{1}$} & \multirow{2}{*}{MCC} \\
    & & & slope  &  & \\
    \midrule
    I  & CCA & 0.52 (0.09) & 3.99 (18.0) & 2.17 (0.64) & 0.06 (0.15) \\
& NI & 0.51 (0.07) & \textbf{1.05} (11.0) & 2.10 (0.55) & 0.05 (0.14) \\
& MI-Wood & 0.50 (0.04) & -0.02 (3.39) & 2.09 (0.53) & 0.02 (0.08) \\
& MI-Bartlett & 0.55 (0.11) & -2.89 (36.7) & 2.17 (0.71) & 0.14 (0.18) \\
& EG & \textbf{0.66} (0.13) & 0.83 (1.02) & \textbf{2.05} (0.82) & \textbf{0.57} (0.06) \\
    II  & CCA & 0.56 (0.12) & \textbf{1.89} (4.70) & 1.96 (0.87) & 0.16 (0.27) \\
& NI & 0.61 (0.14) & 2.64 (2.82) & 1.82 (0.70) & 0.28 (0.31) \\
& MI-Wood & 0.59 (0.13) & 2.40 (3.82) & 1.83 (0.70) & 0.21 (0.27) \\
& MI-Bartlett & 0.64 (0.15) & 6.87 (11.9) & 1.80 (0.72) & 0.35 (0.32) \\
& EG & \textbf{0.77} (0.12) & 2.09 (8.56) & \textbf{1.63} (0.80) & \textbf{0.82} (0.05) \\
    III & CCA & 0.55 (0.11) & 1.83 (3.55) & 1.99 (0.68) & 0.12 (0.23) \\
& NI & 0.54 (0.11) & 3.41 (6.51) & 1.95 (0.60) & 0.13 (0.25) \\
& MI-Wood & 0.52 (0.08) & \textbf{1.20} (1.22) & 1.94 (0.55) & 0.06 (0.18) \\
& MI-Bartlett & 0.57 (0.12) & 3.65 (12.8) & 1.98 (0.63) & 0.25 (0.27) \\
& EG & \textbf{0.67} (0.15) & 4.42 (35.5) & \textbf{1.89} (0.73) & \textbf{0.42} (0.11) \\
    \bottomrule
  \end{tabularx}
  \end{threeparttable}
  \end{center}
\end{table}

\pagebreak

\section{}

A total of $1,762$ NPC patients were included in this analysis with a median follow-up time of $11$ months, in which $266$ of them died due to NPC. 
There were $1,245$ NPC patients whose HPV status was unknown, leaving only 517 with known HPV status, of whom 180 were tested positive for HPV. Web Table \ref{table:desc1} reports descriptive statistics for the study sample of all patients ($n = 1,762$), stratified by whether HPV status was missing or observed. Significant differences were found between these two groups in terms of histologic type ($p < .001$) and AJCC-7 stage ($p = 0.027$). Web Table \ref{table:desc2} presents descriptive statistics for a subgroup of patients with observed HPV status only ($n' = 517$), stratified by HPV+ and HPV-- status. Significant differences between the HPV+ and HPV-- groups were observed for age ($p < .001$), race ($p < .001$), and AJCC-7 M stage ($p = 0.048$). 

\begin{figure}[!ht] 
\centering
\includegraphics[scale=.25]{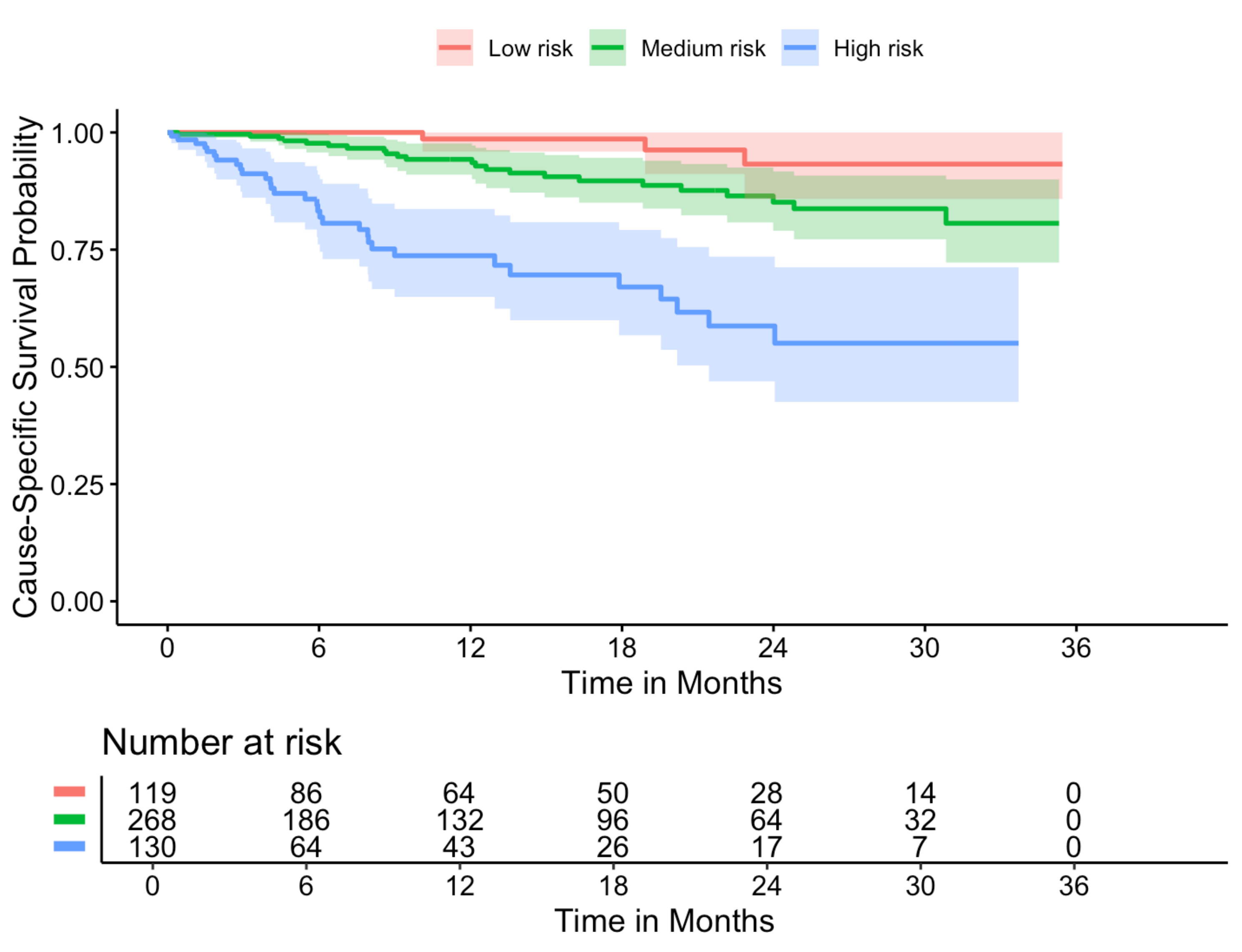}
\caption{Kaplan-Meier curves in the target samples by the three risk groups identified using the expert-guided method for cause-specific survival.}
\label{fig:real}
\end{figure}

Web Figure \ref{fig:real} shows the Kaplan-Meier curves in the target samples by the three risk groups identified using the proposed expert-guided method for cause-specific survival; there was a significant difference in cause-specific survival probabilities across these groups ($p < .001$) based on the log-rank test. The estimated 2-year cause-specific survival (95\% CI) was 94.1\% (87.6\%, 100.0\%) for the low-risk group, 85.1\% (79.2\%, 91.5\%) for the medium-risk group, and 59.6\% (47.6\%, 74.5\%) for the high-risk group. 
A pairwise log-rank test with the Bonferroni-Holm method of adjustment indicated that there were significant pairwise differences between all three groups. \\

\begin{footnotesize}
\begin{longtable}{@{\extracolsep{\fill}} l cccc @{}}
\caption{Descriptive statistics of study samples for all patients stratified by whether HPV status was missing or observed.} 
\label{table:desc1}
\\
\toprule
 & \multicolumn{3}{c}{No. ($\%$) or mean (SD)} & \\
\cmidrule(r){2-4}
Variable & Overall, N = 1762 & Missing, N = 1245 & Observed, N = 517 & $p$-value$^{\textnormal{a}}$ \\
\midrule
\textbf{Gender} &  &  &  &  0.827 \\
Male & 1246 (70.7$\%$)	& 878 (70.5$\%$) & 368 (71.2$\%$) & \\
Female & 516 (29.3$\%$)	& 367 (29.5$\%$) & 149 (28.8$\%$) & \\

\textbf{Age} &  &  &  &  0.051 \\
$<$25  & 73 (4.1$\%$)	& 54 (4.3$\%$)  & 19 (3.7$\%$) & \\
25--49 & 468 (26.6$\%$)	& 330 (26.5$\%$) & 138 (26.7$\%$) & \\
50--74 & 1044 (59.3$\%$)	& 721 (57.9$\%$)  & 323 (62.5$\%$) & \\
75+    & 177 (10.0$\%$)	& 140 (11.2$\%$)  & 37 (7.2$\%$) & \\

\textbf{Martial status} &  &  &  &  0.276 \\
Married & 996 (56.5$\%$) & 698 (56.1$\%$) & 298 (57.6$\%$) & \\
Single  & 380 (21.6$\%$) & 262 (21.0$\%$) & 118 (22.8$\%$) & \\
Others$^{\textnormal{b}}$  & 386 (21.9$\%$)	 & 285 (22.9$\%$) & 101 (19.5$\%$) & \\

\textbf{Race} &  &  &  &  0.062 \\
White & 810 (46.0$\%$)	& 547 (43.9$\%$) & 263 (50.9$\%$) & \\
Black & 212 (12.0$\%$)	& 156 (12.5$\%$) & 56 (10.8$\%$) & \\
East Asian$^{\textnormal{c}}$ & 691 (39.2$\%$)	& 513 (41.2$\%$) & 178 (34.4$\%$) & \\
Others$^{\textnormal{d}}$ & 49 (2.8$\%$)	& 29 (2.3$\%$) & 20 (3.9$\%$) & \\

\textbf{Histologic type} &  &  &  &  $<$.001 \\
Keratinizing$^{\textnormal{e}}$ & 577 (32.7$\%$)	&	373 (30.0$\%$)	& 204 (39.5$\%$)	&\\
Diff/nonkera$^{\textnormal{f}}$ & 434 (24.6$\%$)	&	272 (21.8$\%$)	& 162 (31.3$\%$)	&\\
Undiff/nonkera$^{\textnormal{g}}$ & 251 (14.2$\%$)	&	200 (16.1$\%$)	& 51 (9.9$\%$)	&\\
Others$^{\textnormal{h}}$ & 500 (28.4$\%$)	&	400 (32.1$\%$)	& 100 (19.3$\%$)	& \\

\textbf{AJCC-7 stage} &  &  &  &  0.027 \\
I     & 147 (8.3$\%$) &	105 (8.4$\%$) & 42 (8.1$\%$) &\\
II    & 299 (17.0$\%$) &	204 (16.4$\%$) & 95 (18.4$\%$) &\\
III    & 459 (26.0$\%$) &	304 (24.4$\%$) & 155 (30.0$\%$) &\\
IV$^{\textnormal{i}}$  & 857 (48.6$\%$) & 632 (50.8$\%$) & 225 (43.5$\%$) & \\

\textbf{AJCC-7 T stage} &  &  &  &  0.107 \\
Early stage$^{\textnormal{j}}$    & 1035 (58.7$\%$) &	747 (60.0$\%$) & 288 (55.7$\%$) &\\
Advanced stage$^{\textnormal{k}}$ & 727 (41.3$\%$) & 498 (40.0$\%$) & 229 (44.3$\%$) & \\

\textbf{AJCC-7 N stage} &  &  &  &  0.828 \\
N $=$ 0  & 403 (22.9$\%$) &	287 (23.1$\%$) & 116 (22.4$\%$) &\\
N $>$ 0  & 1359 (77.1$\%$) & 958 (76.9$\%$) & 401 (77.6$\%$) & \\

\textbf{AJCC-7 M stage} &  &  &  &  0.221 \\
M0  & 1572 (89.2$\%$) &	1103 (88.6$\%$) & 469 (90.7$\%$) &\\
M1  & 190 (10.8$\%$) & 142 (11.4$\%$) & 48 (9.3$\%$) & \\

\textbf{Sequence number} &  &  &  &  0.989 \\
One primary only  & 1505 (85.4$\%$) &	1064 (85.5$\%$) & 441 (85.3$\%$) &\\
Others$^{\textnormal{l}}$ & 257 (14.6$\%$) & 181 (14.5$\%$) & 76 (14.7$\%$) & \\

\textbf{Tumor size} & 41.9 (57.2) & 42.7 (61.5) & 40.1 (45.1) & 0.328 \\
\bottomrule
\end{longtable}
\vspace{-1em}
\begin{scriptsize}   
\begin{spacing}{0.5}
\noindent
Abbreviations: Diff/nonkera,  differentiated/nonkeratinizing; Undiff/nonkera = undifferentiated/nonkeratinizing.
$^{\textnormal{a}}$Pearson's Chi-squared test; Welch two-sample t-test.\\
$^{\textnormal{b}}$Others include divorced, separated, unmarried or domestic partner, widowed, and unknown.\\
$^{\textnormal{c}}$East Asian includes Chinese, Japanese, Korean (1988+), and Vietnamese (1988+).\\
$^{\textnormal{d}}$Others include American Indian/Alaska Native, Asian Indian (2010+), Asian Indian or Pakistani-NOS (1988+), Black, Fiji Islander (1991+), Filipino, Guamanian-NOS (1991+), Hawaiian, Hmong (1988+), Kampuchean (1988+), Laotian (1988+), Micronesian-NOS (1991+), Other, Other Asian (1991+), Pacific Islander-NOS (1991+), Pakistani (2010+), Polynesian-NOS (1991+), Samoan (1991+), Thai (1994+), and Tongan (1991+).\\
$^{\textnormal{e}}$Keratinizing squamous cell carcinoma includes 8070 and 8071.\\
$^{\textnormal{f}}$Differentiated non-keratinizing carcinoma includes 8072 and 8073.\\
$^{\textnormal{g}}$Undifferentiated non-keratinizing carcinoma includes 8020, 8021, 8082.\\
$^{\textnormal{h}}$Others include 8000, 8010, 8032, 8041, 8046, 8051, 8074, 8075, 8083,
8090, 8121, 8123, 8140, 8200, 8240, 8246, 8260, 8310, 8430, 8480, 8525,
8560, 8562, 8800, 8801, 8802, 8805, 8890, 8900, 8910, 8920, 8941, 8982,
9364, 9370, 9371, and 9500.\\
$^{\textnormal{i}}$IV includes IVA, IVB, IVC, and IV NOS (Not Otherswise Specified).\\
$^{\textnormal{j}}$Early stage includes T1 and T2.\\
$^{\textnormal{k}}$Advanced stage includes T3, T4, T4a, and T4b.\\
$^{\textnormal{l}}$Others include 1st of 2 or more primaries, 2nd of 2 or more primaries, 3rd
of 3 or more primaries, and 4th of 4 or more primaries.\\
\end{spacing}
\end{scriptsize}
\end{footnotesize}

\pagebreak

\begin{footnotesize}
\begin{longtable}{@{\extracolsep{\fill}} l cccc @{}}
\caption{Descriptive statistics of study samples for a subgroup of patients with observed HPV only stratified by whether a patient was HPV+ or HPV--.} 
\label{table:desc2}
\\
\toprule
 & \multicolumn{3}{c}{No. ($\%$) or mean (SD)} & \\
\cmidrule(r){2-4}
Variable & Overall, N = 517 & HPV+, N = 180 & HPV--, N = 337 & $p$-value$^{\textnormal{a}}$ \\
\midrule
\textbf{Gender} &  &  &  &  0.741 \\
Male & 368 (71.2$\%$) & 126 (70.0\%) & 242 (71.8\%) \\
Female & 149 (28.8$\%$) & 54 (30.0\%) & 95 (28.2\%) \\

\textbf{Age} &  &  &  &  $<$.001 \\
$<$25  & 19 (3.7$\%$) & 12 (6.7\%) & 7 (2.1\%)\\
25--49 & 138 (26.7$\%$) & 39 (21.7\%)	& 99 (29.4\%) \\
50--74 & 323 (62.5$\%$) & 123 (68.3\%) & 200 (59.3\%)	\\
75+    & 37 (7.2$\%$) & 6 (3.3\%) &	31 (9.2\%) \\

\textbf{Martial status} &  &  &  &  0.463 \\
Married & 298 (57.6$\%$) & 98 (54.4$\%$) & 200 (59.3$\%$) & \\
Single  & 118 (22.8$\%$) & 42 (23.3$\%$) & 76 (22.6$\%$) & \\
Others$^{\textnormal{b}}$  & 101 (19.5$\%$) & 40 (22.2$\%$)	 & 61 (18.1$\%$) &\\

\textbf{Race} &  &  &  &  $<$.001 \\
White & 263 (50.9$\%$) & 113 (62.8$\%$)	& 150 (44.5$\%$) & \\
Black & 56 (10.8$\%$) & 18 (10.0$\%$)	& 38 (11.3$\%$) & \\
East Asian$^{\textnormal{c}}$ & 178 (34.4$\%$) & 40 (22.2$\%$)	& 138 (40.9$\%$) & \\
Others$^{\textnormal{d}}$ & 20 (3.9$\%$) & 9 (5.0$\%$)	& 11 (3.3$\%$) & \\

\textbf{Histologic type} &  &  &  &  0.339 \\
Keratinizing$^{\textnormal{e}}$ & 204 (39.5$\%$)	& 78 (43.3$\%$)	&	126 (37.4$\%$)	& \\
Diff/nonkera$^{\textnormal{f}}$ & 162 (31.3$\%$)	& 58 (32.2$\%$)	&	104 (30.9$\%$)	& \\
Undiff/nonkera$^{\textnormal{g}}$ & 51 (9.9$\%$)	& 14 (7.8$\%$)	&	37 (11.0$\%$)	& \\
Others$^{\textnormal{h}}$ & 100 (19.3$\%$)	& 30 (16.7$\%$)	&	70 (20.8$\%$)	& \\

\textbf{AJCC-7 stage} &  &  &  &  0.334 \\
I     & 42 (8.1$\%$) & 10 (5.6$\%$) &	32 (9.5$\%$) & \\
II    & 95 (18.4$\%$) & 38 (21.1$\%$) & 57 (16.9$\%$) & \\
III    & 155 (30.0$\%$) & 53 (29.4$\%$) &	102 (30.3$\%$) & \\
IV$^{\textnormal{i}}$  & 225 (43.5$\%$) & 79 (43.9$\%$) & 146 (43.3$\%$) & \\

\textbf{AJCC-7 T stage} &  &  &  &  0.483 \\
Early stage$^{\textnormal{j}}$    & 288 (55.7$\%$) & 96 (53.3\%)	& 192 (57.0\%)	 \\
Advanced stage$^{\textnormal{k}}$ & 229 (44.3$\%$) & 84 (46.7\%)	& 145 (43.0\%)	 \\

\textbf{AJCC-7 N stage} &  &  &  &  $>$.999 \\
N $=$ 0  & 116 (22.4$\%$) & 40 (22.2\%)	& 76 (22.6\%)	\\
N $>$ 0  & 401 (77.6$\%$) & 140 (77.8\%)	& 261 (77.4\%) \\

\textbf{AJCC-7 M stage} &  &  &  &  0.048 \\
M0  & 469 (90.7$\%$) & 170 (94.4\%)	& 299 (88.7\%)	 \\
M1  & 48 (9.3$\%$) & 10 (5.6\%)	& 38 (11.3\%)	\\

\textbf{Sequence number} &  &  &  &  0.428 \\
One primary only  & 441 (85.3$\%$) & 150 (83.3\%)	& 291 (86.4\%)	 \\
Others$^{\textnormal{l}}$ & 76 (14.7$\%$) & 30 (16.7\%)	& 46 (13.6\%) \\

\textbf{Tumor size} & 40.1 (45.1) & 39.5 (16.2)	& 40.4 (54.6)	& 0.780 \\
\bottomrule
\end{longtable}
\vspace{-1em}
\begin{scriptsize}   
\begin{spacing}{0.5}
\noindent
Abbreviations: Diff/nonkera,  differentiated/nonkeratinizing; Undiff/nonkera = undifferentiated/nonkeratinizing.
$^{\textnormal{a}}$Pearson's Chi-squared test; Welch two-sample t-test.\\
$^{\textnormal{b}}$Others include divorced, separated, unmarried or domestic partner, widowed, and unknown.\\
$^{\textnormal{c}}$East Asian includes Chinese, Japanese, Korean (1988+), and Vietnamese (1988+).\\
$^{\textnormal{d}}$Others include American Indian/Alaska Native, Asian Indian (2010+), Asian Indian or Pakistani-NOS (1988+), Black, Fiji Islander (1991+), Filipino, Guamanian-NOS (1991+), Hawaiian, Hmong (1988+), Kampuchean (1988+), Laotian (1988+), Micronesian-NOS (1991+), Other, Other Asian (1991+), Pacific Islander-NOS (1991+), Pakistani (2010+), Polynesian-NOS (1991+), Samoan (1991+), Thai (1994+), and Tongan (1991+).\\
$^{\textnormal{e}}$Keratinizing squamous cell carcinoma includes 8070 and 8071.\\
$^{\textnormal{f}}$Differentiated non-keratinizing carcinoma includes 8072 and 8073.\\
$^{\textnormal{g}}$Undifferentiated non-keratinizing carcinoma includes 8020, 8021, 8082.\\
$^{\textnormal{h}}$Others include 8000, 8010, 8032, 8041, 8046, 8051, 8074, 8075, 8083,
8090, 8121, 8123, 8140, 8200, 8240, 8246, 8260, 8310, 8430, 8480, 8525,
8560, 8562, 8800, 8801, 8802, 8805, 8890, 8900, 8910, 8920, 8941, 8982,
9364, 9370, 9371, and 9500.\\
$^{\textnormal{i}}$IV includes IVA, IVB, IVC, and IV NOS (Not Otherswise Specified).\\
$^{\textnormal{j}}$Early stage includes T1 and T2.\\
$^{\textnormal{k}}$Advanced stage includes T3, T4, T4a, and T4b.\\
$^{\textnormal{l}}$Others include 1st of 2 or more primaries, 2nd of 2 or more primaries, 3rd
of 3 or more primaries, and 4th of 4 or more primaries.\\
\vspace{1.5em}
\end{spacing}
\end{scriptsize}
\end{footnotesize}

